\documentclass[12pt]{article}

\usepackage{amsmath,graphicx}

\allowdisplaybreaks
\makeatletter 
\newcommand{\mylistoffigs}{\section*{Figure Captions}\@starttoc{fig}}
\newcommand{\figcap}[1]{\refstepcounter{figure}%
  \par\vspace*{1in}\centerline{\textbf{Figure~\thefigure.}} \newpage
  \addcontentsline{fig}{fig}{\noindent\textbf{\thefigure.}~#1}
}
\newcommand{\l@fig}[2]{\par\noindent#1}
\makeatother

\newcommand{\OL}{\overline}

\newcommand{\MAT}[1]{\mathbf{#1}}
\newcommand{\LAB}[1]{\mathsf{#1}}
\newcommand{\BMS}[1]{\boldsymbol#1}
\newcommand{\BMSS}[1]{\boldsymbol#1}
\newcommand{\FRAC}[2]{\textstyle\frac{#1}{#2}}
\newcommand{\KET}[1]{|#1\rangle}
\newcommand{\BRA}[1]{\langle#1|}
\newcommand{\HALF}{{\FRAC{1}{2}}}
\newcommand{\UPS}{{\BMS\iota}}		
\newcommand{\UPSS}{{\BMSS\iota}}	
\newcommand{\RHO}{{\BMS\rho}}

\newcommand{\EP}{\MAT{E}_{+}}
\newcommand{\EM}{\MAT{E}_{-}}

\newcommand{\EPM}{\MAT{E}_{\pm}}

\newcommand{\IX}{\MAT{I}_\LAB{x}}
\newcommand{\IY}{\MAT{I}_\LAB{y}}
\newcommand{\IZ}{\MAT{I}_\LAB{z}}

\newenvironment{Desc} {\begin{list}{}{%
\setlength{\labelwidth}{-6pt}%
\setlength{\leftmargin}{0pt}}}%
{\end{list}}

\title{
\vspace*{1in} 
A Study of Quantum Error Correction by Geometric
Algebra and Liquid-State NMR Spectroscopy
\vspace*{0.25in} 
}
\author{
Y.\ Sharf and D.\ G.\ Cory \\
Dept.\ of Nuclear Engineering \\
Massachusetts Institute of Technology \\
Cambridge, MA 02139 \\
\vspace*{0.25in} 
\\ \and
S.\ S.\ Somaroo and T.\ F.\ Havel\footnote{
To whom correspondence should be addressed.} \\
BCMP, Harvard Medical School \\
240 Longwood Ave., Boston, MA 02115 \\
\vspace*{0.25in} 
\\ \and
E.\ Knill (CIC-3), R.\ Laflamme \\
and W.\ H.\ Zurek (T-6) \\
Los Alamos National Laboratory \\
Los Alamos, NM 87545
}
\date{%
\vspace*{0.5in}
\today}

\begin{document} \maketitle 

\pagebreak 
\renewcommand{\baselinestretch}{1.5} 

\begin{abstract}
Quantum error correcting codes enable the information
contained in a quantum state to be protected
from decoherence due to external perturbations.
Applied to NMR, quantum coding does not alter normal relaxation,
but rather converts the state of a ``data'' spin into multiple
quantum coherences involving additional {\em ancilla\/} spins.
These multiple quantum coherences relax at differing rates,
thus permitting the original state of the data to be approximately
reconstructed by mixing them together in an appropriate fashion.
This paper describes the operation of a simple,
three-bit quantum code in the product operator formalism,
and uses geometric algebra methods to obtain the
error-corrected decay curve in the presence of
arbitrary correlations in the external random fields.
These predictions are confirmed in both the totally
correlated and uncorrelated cases by liquid-state NMR
experiments on ${}^{13}\LAB{C}$-labeled alanine,
using gradient-diffusion methods to implement
these idealized decoherence models.
Quantum error correction in weakly polarized systems
requires that the ancilla spins be prepared in a
{\em pseudo-pure\/} state relative to the data spin,
which entails a loss of signal that exceeds
any potential gain through error correction.
Nevertheless, this study shows that quantum coding
can be used to validate theoretical decoherence
mechanisms, and to provide detailed information on
correlations in the underlying NMR relaxation dynamics.
\end{abstract}

\renewcommand{\baselinestretch}{1.5}
\setlength{\jot}{9pt}	
\renewcommand{\floatpagefraction}{0}
\renewcommand{\arraystretch}{1.5}
\renewcommand{\baselinestretch}{1.5}
\normalsize	
\pagebreak \setcounter{page}{1} 

\section{Introduction} \label{sec:int}
A quantum computer stores binary information
in an array of two-state quantum systems,
e.g.\ spin $\FRAC{1}{2}$ nuclei.
It performs logical operations on this information
via unitary transformations obtained by controlling
the effective interactions among the systems.
By operating on a coherent superposition over all
combinations of states of the individual systems,
it can, in effect, operate on all these combinations at once.
This yields a degree of parallelism which grows
exponentially with the size of the problem.
Because such superpositions are extremely sensitive to
decoherence, it was initially doubted that they could be
maintained long enough to perform significant computations.
Shor \cite{Shor:95} and Steane \cite{Steane:96}
were nonetheless able to devise error correcting
codes to control decoherence in quantum computers.
It has now been shown that arbitrarily long quantum
computations can be carried out providing the
error rate per operation is below some threshold
\cite{KnillLafla:97,Gottesman:98,Steane:98b,Preskill:99}.

Even though a true quantum computer has yet to be built,
small quantum computations can be performed using
standard liquid-state NMR spectroscopy to operate
on the nuclear spins in ensembles of molecules in
pseudo-pure states \cite{CorFahHav:97,GershChuan:97}.
These states may be characterized as having a density
matrix with only a single nondegenerate eigenvalue,
whose corresponding eigenvector transforms
identically to the state vector of a true pure state.
This approach has enabled all the basic operations of quantum
computing to be demonstrated \cite{CorPriHav:98,ChGeKuLe:98},
including simple quantum algorithms
\cite{JonMosHan:98,LinBarFre:98,ChVaZhLeLl:98},
teleportation \cite{NieKniLaf:98},
and error correction \cite{CMPKLZHS:98}.
The preparation of pseudo-pure states from
equilibrium states nevertheless entails a
rapid loss in signal with increasing numbers
of spins \cite{HaSoTsCo:99,KniChuLaf:98,Warren:97},
precluding the use of liquid-state NMR as a means
of performing large-scale quantum computations.

This paper builds on the authors' earlier implementation of
a three-bit quantum error correcting code \cite{CMPKLZHS:98},
and consists of two parts, one theoretical and the other experimental.
In the first part, geometric algebra methods \cite{SomCorHav:98}
are used to derive the error corrected decay curve assuming
arbitrary correlations in the relaxation processes responsible.
The second describes a complete NMR implementation of the code,
and demonstrates that it performs as predicted by the
theory using gradient diffusion methods to implement both
totally correlated and uncorrelated $T_2$ relaxation.
The theoretical (sections {\bf \ref{sec:enc} -- \ref{sec:dec}})
and experimental (sections {\bf \ref{sec:grd} -- \ref{sec:res}})
parts of the paper may be read largely independently of each other.
Section {\bf \ref{sec:qec}} provides an overview
of the three-bit quantum error correcting code,
while section {\bf \ref{sec:dis}} discusses the
significance of the theory in the context of NMR.
Together, the results of this paper show how quantum
coding can be used not only to correct for decoherence,
but also to design experiments to test specific
hypotheses on the underlying relaxation mechanisms.

\section{The Quantum Error Correcting Code} \label{sec:qec}
The theory of quantum error correcting codes,
though only a few years old \cite{Shor:95,Steane:96},
is already a well-established and highly developed subject
\cite{KnillLafla:97,Gottesman:98,Steane:98b,Preskill:99}.
The ``majority logic'' code demonstrated in this paper
protects information encoded in the joint state
of three spins against arbitrary spin flip errors.
Given a data spin (throughout this paper, the first)
in some arbitrary state $\alpha\KET{0} + \beta\KET{1}$
(where $\KET{0}$ is the ``down'' (ground) state of the spin,
$\KET{1}$ the ``up'' state, and $|\alpha|^2 + |\beta|^2 = 1$),
two ancilla spins in their ground state are added to the system,
obtaining $(\alpha\KET{0} + \beta\KET{1}) \KET{00}$.
A unitary transformation is then applied to the system
consisting of two ``c-NOT'' (controlled-NOT) gates
conditional on the data spin (see Fig.\ \ref{fig:diagram}).
This step, called encoding, produces the state
$\alpha\KET{000} + \beta\KET{111}$.

Note that rotating a spin by $\pi$ in this state
is the same as interchanging a pair of corresponding
bits between $\KET{000}$ and $\KET{111}$,
and results in an orthogonal state.
Thus one can determine if a single spin flip has occured,
and which spin it happened to, by a measurement 
that reveals which of the orthogonal subspaces
$\alpha \KET{100} + \beta \KET{011}$,
$\alpha \KET{010} + \beta \KET{101}$,
$\alpha \KET{001} + \beta \KET{110}$
(if any) the system has moved into.
If no error occured, this measurement
leaves the state of the system intact,
and otherwise one knows which spin needs to be
rotated by $\pi$ in order to fix the error.
This illustrates the basic idea behind
all quantum error correcting codes:
the correctable errors map the subspace of
valid states onto mutually orthogonal subspaces,
in each of which the unitary operation
needed to fix the error is known.
Because these entangled subspaces cannot be measured
directly, the state must be decoded so that the
error syndrome is contained in the ancillae alone.
This is done with the same unitary transformation
as encoding, since it is its own inverse.

In practice, the spins are subject to numerous small
random perturbations from the environment arising,
for example, from fluctuating external random fields,
rather than single spin flips as above.
Throughout this paper, these fields will be assumed to be
about the $\LAB{x}$-axis in the rotating frame \cite{Slichter:90}.
To the extent that these perturbations affect
multiple spins, the data spin will lose coherence
in the $\LAB{yz}$-plane despite error correction.
It nevertheless follows from the linearity
of quantum mechanics that the code will still
prevent decoherence to first order in time,
as will now be shown.

The interaction of such fields with the $\LAB{y}$ and
$\LAB{z}$-components of the spins has the Hamiltonian
\begin{equation}
\MAT H_\LAB{x} ~=~ \gamma^1 B_\LAB{x}^1(t) \IX^1 +
\gamma^2 B_\LAB{x}^2(t) \IX^2 + \gamma^3 B_\LAB{x}^3(t) \IX^3 ~,
\end{equation}
where $B_\LAB{x}^k$ denotes the $\LAB{x}$-component
of the fluctuating magnetic field at each spin,
$\gamma^k$ is its gyromagnetic ratio, and $\IX^k
\equiv \HALF\BMS\sigma_\LAB{x}^k$ are the usual angular
momentum operators in units of $\hbar$ ($k = 1,2,3$).
The corresponding propagator has the form
$\exp(-\UPS(\chi^1 \IX^1 + \chi^2 \IX^2 + \chi^3 \IX^3))$,
where $\UPS$ denotes the imaginary unit and
$\chi^k = \gamma^k \int_0^t B_\LAB{x}^k(\tau) {\rm d}\tau$.
In the limit of short decoherence times and
hence small accumulated phase errors $\chi^k$,
a first-order expansion of this propagator is adequate,
namely:
\begin{equation}
\exp\left( -\UPS (\chi^1 \IX^1 + \chi^2 \IX^2 + \chi^3 \IX^3 ) \right)
~\approx~ 1 - \UPS \chi^1 \IX^1 - \UPS \chi^2 \IX^2 - \UPS \chi^3 \IX^3
\end{equation}
Thus to first order this random propagator maps the encoded state to
\begin{equation} \begin{array}{rcl}
(\alpha\KET{000} + \beta\KET{111}) &-&
\FRAC{\UPS}{2\,}\chi^1 (\alpha\KET{100} + \beta\KET{011}) \\ &-&
\FRAC{\UPS}{2\,}\chi^2 (\alpha\KET{010} + \beta\KET{101}) \\ &-&
\FRAC{\UPS}{2\,}\chi^3 (\alpha\KET{001} + \beta\KET{110}) ~.
\end{array} \end{equation}
The decoding operation maps this to
\begin{equation} \begin{array}{rcl}
(\alpha\KET{0} + \beta\KET{1}) \KET{00} &-&
\FRAC{\UPS\,}{2\,}\chi^1 (\alpha\KET{1} + \beta\KET{0}) \KET{11} \\ &-&
\FRAC{\UPS\,}{2\,}\chi^2 (\alpha\KET{0} + \beta\KET{1}) \KET{10} \\ &-&
\FRAC{\UPS\,}{2\,}\chi^3 (\alpha\KET{0} + \beta\KET{1}) \KET{01} ~.
\end{array} \end{equation}
Finally, a Toffoli gate (also called a ``c${}^2$-NOT'')
corrects the error, by flipping the data spin in the
term in which both ancillae are ``down'' ($\KET{11}$).
This returns the data spin to its original state
$\alpha\KET{0} + \beta\KET{1}$ in that term,
and otherwise leaves it alone.

Thus, in addition to single spin flips, the code
corrects to first order for random, time-dependent
rotations about the $\LAB{x}$-axis as claimed.
By inserting a $\pi/2$ $\LAB{y}$-rotation of all the
spins after encoding and its inverse before decoding,
this code can be transformed to one that protects
against random $\LAB{z}$-rotations, which give rise
to adiabatic $T_2$ relaxation in NMR \cite{CMPKLZHS:98}.
Because the code does {\em not\/} protect the ancillae,
they must be ``discarded'' after correction, which
corresponds to taking the partial trace over them.
In NMR spectroscopy, this operation is performed by
{\em decoupling\/} the ancillae from the data during acquisition.
Note also that, since strong measurements are not
available in ensembles, it is not possible to reset
and reuse the ancillae for repeated error correction
(see section {\bf \ref{sec:dis}} for further discussion of this issue).

The following three sections will present
a detailed theoretical analysis of the code
in the product operator notation of NMR
\cite{Slichter:90,ErnBodWok:87}, using
geometric algebra methods \cite{SomCorHav:98}
to compute the results of the encoding,
decoherence, decoding and correction operations.
Readers who are interested primarily in the
experimental procedures and results should
proceed directly to the main theoretical
result, Eq.\ (\ref{eq:final_result}),
and continue reading from there.

\section{Geometric Algebra Analysis of Encoding} \label{sec:enc}
In NMR spectroscopy, one does not observe
individual spin systems, but rather averages
over macroscopic ensembles of such systems.
In the semi-classical treatment of liquid-state
NMR relaxation processes used here \cite{Abragam:61},
two distinct types of averages must be taken.
The first is an average over the initial state
of a typical molecule, which results, for example,
in the equilibrium state $\IZ^1 + \IZ^2 + \IZ^3$.
The second is an average over the environment
of a typical molecule, or more precisely,
the external random fields which act upon it.
Provided the random fields are uncorrelated with the
initial state, these two averages can be taken separately.
In order to focus on how the results of error
correction are averaged over the environment,
in the following three sections it will be assumed
that the spin system is initially in a pure state;
the effect of averaging over an ensemble of spin systems
in different pure states is considered in an Appendix.

Even though an initial pure state is assumed,
the random fields from the environment will cause
the state to decohere into a general mixed state,
which must be described by a density matrix.
The physical meaning of a density matrix is most
clearly expressed using the product operator formalism,
since it is based on a complete system of observables
\cite{BoulaRance:94a,SoEiLeBoEr:83,vdVenHilbe:83}.
In the algebra generated by the product operators,
spinors (pure states) correspond to {\em idempotents\/},
namely elements that are equal to their squares.
In particular, spinors of the form
$\KET{\delta^1\delta^2\delta^3}$
($\delta^k = 0, 1$; $k = 1,2,3$)
correspond to products of three idempotents,
namely:
\begin{equation}
\EPM^1\EPM^2\EPM^3 ~\equiv~
\HALF (1 + (-1)^{\delta^1} 2\IZ^1)
\HALF (1 + (-1)^{\delta^2} 2\IZ^2)
\HALF (1 + (-1)^{\delta^3} 2\IZ^3)
~\longleftrightarrow~ 
\KET{\delta^1\delta^2\delta^3}
\BRA{\delta^1\delta^2\delta^3}
\end{equation}
(Note that throughout this paper the scalar identity $1$ is
identified with the $3$-spin identity operator ${\bf 1}_8$.)
Letting $\tilde\alpha, \tilde\beta$ be the
complex conjugates of $\alpha, \beta \in \cal C$,
and $\Re$, $\Im$ denote the real and imaginary parts,
the density matrix of the data spin \#$1$ in a general
superposition state $\alpha\KET{0} + \beta\KET{1}$ is:
\begin{eqnarray} \label{eq:init}
\RHO_\LAB{A}^1 &\equiv& (\alpha\KET{0} + \beta\KET{1})
(\tilde\alpha\BRA{0} + \tilde\beta\BRA{1})
\nonumber \\ &=&
\HALF(|\alpha|^2 + |\beta|^2) (\KET{0}\BRA{0} + \KET{1}\BRA{1})
+ \Re(\tilde\alpha\beta) (\KET{1}\BRA{0} + \KET{0}\BRA{1}) \,+
\nonumber \\ &&
\UPS\Im(\tilde\alpha\beta)(\KET{1}\BRA{0} - \KET{0}\BRA{1}) 
+ \HALF(|\alpha|^2 - |\beta|^2) (\KET{0}\BRA{0} - \KET{1}\BRA{1})
\\ \nonumber &=& \HALF(|\alpha|^2 + |\beta|^2) +
\Re(\tilde\alpha\beta) \, 2\IX^1 + \Im(\tilde\alpha\beta)
\, 2\IY^1 + (|\alpha|^2 - |\beta|^2) \, \IZ^1
\\ \nonumber &=&
(\alpha + 2 \beta \IX^1) \EP^1 (\tilde\alpha + 2 \tilde\beta \IX^1)
\end{eqnarray}
Alternatively, an arbitrary state can be obtained by
rotation of the $\KET{0}\BRA{0} \equiv \EP^1$ state,
which may be specified in terms of polar angles as
\begin{eqnarray}
\RHO_\LAB{A}^1 &\equiv&
e^{-\UPSS\phi\IZ^1} e^{-\UPSS\theta\IX^1}
\EP^1 e^{\UPSS\theta\IX^1} e^{\UPSS\phi\IZ^1}
\nonumber \\ &=&
e^{-\UPSS\phi\IZ^1} (\cos(\FRAC{\theta}{2})
- 2\UPS\sin(\FRAC{\theta}{2})\IX^1) \EP^1
(\cos(\FRAC{\theta}{2}) +
2\UPS\sin(\FRAC{\theta}{2})\IX^1)
e^{\UPSS\phi\IZ^1}
\\ \nonumber &=&
(\cos(\FRAC{\theta}{2}) e^{-\UPSS\phi\IZ^1}
- 2\UPS\sin(\FRAC{\theta}{2})\IX^1
e^{\UPSS\phi\IZ^1}) \EP^1
(e^{\UPSS\phi\IZ^1}
\cos(\FRAC{\theta}{2}) + e^{-\UPSS\phi\IZ^1}
2\UPS\sin(\FRAC{\theta}{2})\IX^1)
\\ \nonumber &=&
(\cos(\FRAC{\theta}{2}) e^{-\UPSS\phi/2}
- 2\UPS\sin(\FRAC{\theta}{2})\IX^1
e^{\UPSS\phi/2}) \EP^1 (e^{\UPSS\phi/2}
\cos(\FRAC{\theta}{2}) + e^{-\UPSS\phi/2}
2\UPS\sin(\FRAC{\theta}{2})\IX^1) ~,
\end{eqnarray}
where the last line was obtained by using
the relation $\IZ^1 \EPM^1 = \pm\HALF \EPM^1$
to absorb the $\IZ^1$ operators in the exponentials.
A comparison of these two expressions for
$\RHO_\LAB{A}^1$ shows the coefficients in the
superposition may be interpreted geometrically as
\begin{equation}
\alpha ~\equiv~ \cos(\FRAC{\theta}{2}) \, e^{-\UPSS\phi/2} ~, \qquad
\beta ~\equiv~ -\UPS \sin(\FRAC{\theta}{2}) \, e^{\UPSS\phi/2} ~.
\end{equation}

This correspondence between states and operators,
which is the basis of the product operator formalism,
has been used together with reformulations of multiparticle
quantum mechanics in the language of geometric algebra
\cite{DorLasGul:93,SomLasDor:99,DoLaGuSoCh:96,Hestenes:66}
to obtain algebraic expressions and geometric interpretations
for the basic operations of quantum computing \cite{SomCorHav:98}.
Of particular interest is the fact that the propagators of
c-NOT gates can also be written in terms of idempotents.
Thus the particular c-NOT gate that flips the
second spin when the first is ``up'' has the form:
\begin{eqnarray}
\MAT{S}^{2|1} &\equiv& \exp(\UPS\pi\EM^1(1 - 2\IX^2)/2)
~=~ \EM^1 \, \exp(\UPS\pi(1 - 2\IX^2)/2) + (1 - \EM^1)
\\ \nonumber
&=& 2 \IX^2 \EM^1 + \EP^1 ~=~ 1 - \EM^1(1 - 2\IX^2)
\end{eqnarray}
Using the relations $4 \IX^2 \IX^2 = 1$,
$\IX^2 \EP^2 = \EM^2 \IX^2$ and
$\EP^1\EM^1 = \EM^1\EP^1 = 0$,
$\MAT{S}^{2|1}$ may be shown to exchange $\EP^2$
with $\EM^2$ in those products containing $\EM^1$;
for example:
\begin{eqnarray}
\MAT{S}^{2|1} (\EM^1 \EP^2) \MAT{S}^{2|1} &=&
(2\IX^2 \EM^1) (\EM^1 \EP^2) (\EM^1 2 \IX^2) +
\EP^1 (\EM^1 \EP^1) \EP^1 \nonumber \\
&=& \EM^1 (2\IX^2 \HALF (1 + 2 \IZ^2)\, 2\IX^2) \\
&=& \EM^1 \HALF (1 - 8 \IZ^2 \IX^2 \IX^2)
~=~ \EM^1 \EM^2 \nonumber
\end{eqnarray}
In a similar fashion, it can be shown that
the Toffoli gate, which flips the first spin
only if the other two are ``up'', has the form:
\begin{eqnarray}
\MAT T^{1|23} &\equiv& \exp(\UPS\pi(1 - 2\IX^1)\EM^2\EM^3)
\\ \nonumber
&=& 2\IX^1\EM^2\EM^3 + (1 - \EM^2\EM^3)
~=~ 1 - (1 - 2\IX^1) \EM^2\EM^3
\end{eqnarray}
These are all the gates that are needed to implement
the error correction procedure studied in this paper,
whose operation will now be analyzed using geometric algebra.

The first step of the procedure is to encode
the state $\RHO_\LAB{A}^1$ of the first spin by
correlating it with the ancillae in the state $\EP^2 \EP^3$.
This is done by applying the c-NOT's $\MAT{S}^{2|1}$ and
$\MAT{S}^{3|1}$, as shown in Fig.\ \ref{fig:diagram}.
These c-NOT's commute, and their product is
\begin{eqnarray}
\MAT{S}^{2|1} \MAT{S}^{3|1} &=&
(2 \IX^2 \EM^1 + \EP^1) (2 \IX^3 \EM^1 + \EP^1)
\\ &=& \nonumber
4 \IX^2 \IX^3 \EM^1 + \EP^1
~\equiv~ \MAT{S}^{23|1} ~.
\end{eqnarray}
It follows that
\begin{eqnarray}
\MAT{S}^{23|1} 2\IX^1 \MAT{S}^{23|1}
&=& (4 \IX^2 \IX^3 \EM^1 + \EP^1) (2 \IX^1) (4 \IX^2 \IX^3 \EM^1 + \EP^1)
\\ \nonumber &=&
(2 \IX^1) (4 \IX^2 \IX^3 \EP^1 + \EM^1) (4 \IX^2 \IX^3 \EM^1 + \EP^1)
\\ \nonumber &=&
(2 \IX^1) (4 \IX^2 \IX^3 (\EP^1 + \EM^1)) ~=~ 8 \IX^1 \IX^2 \IX^3 ~.
\end{eqnarray}
Since $\MAT{S}^{23|1} \MAT{S}^{23|1} = 1$ and
$\MAT{S}^{23|1}$ commutes with $\EP^1\EP^2\EP^3$,
the density matrix $\RHO_\LAB{B}$ of the encoded state,
which is obtained by applying $\MAT{S}^{23|1}$ to the initial density
matrix $\RHO_\LAB{A} \equiv \RHO_\LAB{A}^1 \EP^2 \EP^3$, is given by
\begin{eqnarray}
\RHO_\LAB{B} &\equiv&
\MAT{S}^{23|1} \RHO_\LAB{A} \MAT{S}^{23|1}
\nonumber \\ &=&
\MAT{S}^{23|1} (\alpha + \beta 2\IX^1)\, \MAT{S}^{23|1}
\MAT{S}^{23|1} (\EP^1\EP^2\EP^3) \MAT{S}^{23|1}
\MAT{S}^{23|1}  (\tilde\alpha + \tilde\beta 2\IX^1) \, \MAT{S}^{23|1}
\\ &=& \nonumber
(\alpha + \beta 8 \IX^1 \IX^2 \IX^3)
(\EP^1\EP^2\EP^3) (\tilde\alpha + \tilde\beta 8 \IX^1 \IX^2 \IX^3) ~.
\end{eqnarray}
The product of all three $\IX$ operators in this expression
reflects the entanglement of the three spins in the encoded state.

\section{Averaging Over the Environment} \label{sec:avg}
The next step is to apply a random rotation
about the $\LAB{x}$-axis to the encoded state,
and then to compute the average of the result.
Fortunately, the $\LAB{x}$-rotation commutes
with the outer factors of $\RHO_\LAB{B}$, so that
it can be applied directly to the inner factor:
\begin{eqnarray} \label{eq:rhoB}
&& e^{-\UPSS(\chi^1\IX^1 + \chi^2\IX^2 + \chi^3\IX^3)}
\RHO_\LAB{B}
e^{\UPSS(\chi^1\IX^1 + \chi^2\IX^2 + \chi^3\IX^3)}
\\ &=& \nonumber
(\alpha + \beta 8 \IX^1 \IX^2 \IX^3)
e^{-\UPSS(\chi^1\IX^1 + \chi^2\IX^2 + \chi^3\IX^3)}
(\EP^1\EP^2\EP^3)
e^{\UPSS(\chi^1\IX^1 + \chi^2\IX^2 + \chi^3\IX^3)}
(\tilde\alpha + \tilde\beta 8 \IX^1 \IX^2 \IX^3)
\end{eqnarray}
Since the outer factors are constant,
they can be taken out of the ensemble average.
This reduces the analysis of the decoherence process
for arbitrary $\alpha$ and $\beta$ to the special
case in which $\alpha = 1$ and $\beta = 0$.

To facilitate this process,
rewrite the inner transformation as
\begin{eqnarray} \label{eq:product}
&& \left( e^{-\UPSS\chi^1\IX^1} \EP^1
e^{\UPSS\chi^1\IX^1} \right)
\left( e^{-\UPSS\chi^2\IX^2} \EP^2
e^{\UPSS\chi^2\IX^2} \right)
\left( e^{-\UPSS\chi^3\IX^3} \EP^3
e^{\UPSS\chi^3\IX^3} \right)
\nonumber \\ &=& \FRAC{1}{8}
\left( 1 + e^{-\UPSS\chi^1\IX^1}
2\IZ^1 e^{\UPSS\chi^1\IX^1} \right)
\left( 1 + e^{-\UPSS\chi^2\IX^2}
2\IZ^2 e^{\UPSS\chi^2\IX^2} \right)
\left( 1 + e^{-\UPSS\chi^3\IX^3}
2\IZ^3 e^{\UPSS\chi^3\IX^3} \right)
\\ &=& \FRAC{1}{8} \nonumber
\left( 1 + e^{-2\UPSS\chi^1\IX^1} 2\IZ^1 \right)
\left( 1 + e^{-2\UPSS\chi^2\IX^2} 2\IZ^2 \right)
\left( 1 + e^{-2\UPSS\chi^3\IX^3} 2\IZ^3 \right) ~.
\end{eqnarray}
Expanding this product yields the sum
of all possible terms of the form
\begin{equation}
\MAT{R}_{\delta^1\delta^2\delta^3}(\chi^1,\chi^2,\chi^3)
\left( (1 - \delta^1) + \delta^1 2\IZ^1 \right)
\left( (1 - \delta^2) + \delta^2 2\IZ^2 \right)
\left( (1 - \delta^3) + \delta^3 2\IZ^3 \right)
\end{equation}
where
\begin{equation}
\MAT{R}_{\delta^1\delta^2\delta^3}(\chi^1,\chi^2,\chi^3)
~\equiv~ e^{-2\UPSS( \delta^1\chi^1\IX^1 +
\delta^2\chi^2\IX^2 + \delta^3\chi^3\IX^3 )}
\end{equation}
for $\delta^1,\delta^2,\delta^3 \in \{0,1\}$.

It follows that the average state can be
obtained by applying the averages of the propagators
$\OL{\MAT{R}_{\delta^1\delta^2\delta^3}(\chi^1,\chi^2,\chi^3)}$
to the corresponding terms of the encoded state $\RHO_\LAB{B}$.
Assuming that the spectral density of the random fields
$B_\LAB{x}(t)$ is well-approximat\-ed by a delta-function
(i.e., that the correlation time is small compared to
the inverse square root of the variance in the frequencies
$\gamma B_\LAB{x}(t)$; see \cite{Cowan:97} for details),
the joint probability density function of the random
angles $\chi^1, \chi^2, \chi^3$ will be a multivariate
Gaussian whose covariance matrix grows linearly in time.
Letting $\BMS\chi \equiv [\chi^1,\chi^2,\chi^3]^\top$ and
$[\OL{\chi^k(t) \chi^\ell(t)}] \equiv [c^{jk} t] = \MAT{C}\,t$
be the covariance matrix, this can be written as
\begin{equation} \label{eq:cov_def}
P(\BMS\chi) ~=~ \left( (2\pi)^3\,\det(\MAT{C}\,t) \right)^{-\frac{1}{2}}
e^{-\frac{1}{2t} \BMSS{\chi}^\top \MAT{C}^{-1} \BMSS{\chi}} ~.
\end{equation}

At this point it turns out to be easier to
rotate $\MAT{R}_{\delta^1\delta^2\delta^3}%
(\chi^1,\chi^2,\chi^3)$ to the $\LAB{z}$-axis:
\begin{eqnarray}
\MAT{R}_{\delta^1\delta^2\delta^3}'(\chi^1,\chi^2,\chi^3)
&\equiv& e^{\UPSS\frac{\pi}{2}(\IY^1 + \IY^2 + \IY^3)}
\MAT{R}_{\delta^1\delta^2\delta^3}(\chi^1,\chi^2,\chi^3)
e^{-\UPSS\frac{\pi}{2}(\IY^1 + \IY^2 + \IY^3)}
\nonumber \\
&=& e^{-2\UPSS( \delta^1\chi^1\IZ^1 +
\delta^2\chi^2\IZ^2 + \delta^3\chi^3\IZ^3 )}
\end{eqnarray}
This is because the idempotents $\EPM$
``absorb'' the $\IZ$ operators in the exponent,
so that it can be expanded into a sum of terms
involving only {\em scalar\/} exponentials,
\begin{equation}
\MAT{R}_{\delta^1\delta^2\delta^3}'(\chi^1,\chi^2,\chi^3) ~=~
\MAT{R}_{\delta^1\delta^2\delta^3}'(\chi^1,\chi^2,\chi^3)
(\EP^1 + \EM^1) (\EP^2 + \EM^2) (\EP^3 + \EM^3)
\end{equation}
where each term has the form
\begin{equation}
\MAT{R}_{\delta^1\delta^2\delta^3}'(\chi^1,\chi^2,\chi^3)
\MAT{E}_{\epsilon^1}^1 \MAT{E}_{\epsilon^2}^2 \MAT{E}_{\epsilon^3}^3
~\equiv~
e^{-\UPSS(\epsilon^1\chi^1 + \epsilon^2\chi^2 + \epsilon^3\chi^3)}
\MAT{E}_{\epsilon^1}^1 \MAT{E}_{\epsilon^2}^2 \MAT{E}_{\epsilon^3}^3
\end{equation}
with $\epsilon^k \in \{-1,0,+1\}$, $\epsilon^k = \delta^k
\epsilon^k$ and $\MAT{E}_0^k \equiv 1$ for $k = 1,2,3$.
The average of each such scalar exponential is
\begin{equation} \label{eq:scalar_integ}
\OL{e^{-\UPSS(\epsilon^1\chi^1 + \epsilon^2\chi^2 + \epsilon^3\chi^3)}} ~=~
\int_{-\infty}^{\infty} P(\BMS\chi) e^{-\UPSS\BMSS\chi\cdot\BMSS\epsilon}
{\rm d}\BMS\chi ~=~ e^{-\frac{t}{2} \BMSS\epsilon^\top \MAT{C} \BMSS\epsilon} ~,
\end{equation}
where $\BMS\epsilon \equiv [\epsilon^1,\epsilon^2,\epsilon^3]^\top$.
Since this is independent of the overall sign of $\BMS\epsilon$,
for $\BMS\epsilon \ne \BMS 0$
the ensemble average of each $\MAT{R}_{\delta^1\delta^2\delta^3}'%
(\chi^1,\chi^2,\chi^3)$ is a sum of pairs of terms of the form
\begin{eqnarray} \label{eq:pair_to_aver}
e^{-\frac{t}{2} \BMSS\epsilon^\top \MAT{C} \BMSS\epsilon}
\left( \MAT{E}_{\epsilon^1}^1 \MAT{E}_{\epsilon^2}^2 \MAT{E}_{\epsilon^3}^3
+ \MAT{E}_{-\epsilon^1}^1 \MAT{E}_{-\epsilon^2}^2 \MAT{E}_{-\epsilon^3}^3
\right) ~.
\end{eqnarray}

In the case that $\epsilon^1 = \epsilon^2 = \epsilon^3 = 0$,
Eq.\ (\ref{eq:scalar_integ}) yields $1$ as expected,
while in the case that $\epsilon^j = \pm 1$ while the
other two $\epsilon^i = 0$ ($1 \le i \ne j \le 3$),
the two terms in Eq.\ (\ref{eq:pair_to_aver}) are
\begin{equation}
\OL{e^{-2 \UPSS \chi^j \IZ^j}} ~=~ e^{-\frac{t}{2} c^{jj}}
(\EP^j + \EM^j) ~=~ e^{-t c^{jj} / 2} ~.
\end{equation}
In the case that $\epsilon^i = 0$ while $\epsilon^j$
and $\epsilon^k$ are nonzero ($\{i,j,k\} = \{1,2,3\}$),
Eq.\ (\ref{eq:pair_to_aver}) has the form
\begin{equation}
e^{-\frac{t}{2}(c^{jj} + c^{kk} + 2 \epsilon^j \epsilon^k c^{jk})}
(\MAT{E}_{\epsilon^j} \MAT{E}_{\epsilon^k} +
\MAT{E}_{-\epsilon^j} \MAT{E}_{-\epsilon^k})
\end{equation}
where
\begin{equation}
(\MAT{E}_{\epsilon^j} \MAT{E}_{\epsilon^k} +
\MAT{E}_{-\epsilon^j} \MAT{E}_{-\epsilon^k})
~=~ \HALF (1 + 4 \epsilon^j \epsilon^k \IZ^j \IZ^k)
~\equiv~ \MAT{E}_{\epsilon^j\epsilon^k}^{jk}
\end{equation}
and $\EPM^{jk} \equiv \HALF (1 \pm 4\IZ^j\IZ^k)$ is idempotent.
Since there are two such pairs of terms, one with $\epsilon^j = \epsilon^k$
and the other with $\epsilon^j = -\epsilon^k$, the complete result is
\begin{eqnarray}
\OL{e^{-2\UPSS (\chi^j \IZ^j + \chi^k \IZ^k)}}
&=& e^{-\frac{t}{2}(c^{jj} + c^{kk})} \left(
e^{-t\,c^{jk}} \EP^{jk} + e^{t\,c^{jk}} \EM^{jk} \right)
\nonumber \\ &=&
e^{-\frac{t}{2}(c^{jj} + c^{kk})}
\left( \cosh(t\,c^{jk}) - \sinh(t\,c^{jk})\, 4\IZ^j\IZ^k \right)
\\ &=& \nonumber
e^{-\frac{t}{2} (c^{jj} + c^{kk} + 8 c^{jk} \IZ^j \IZ^k)} ~.
\end{eqnarray}
In a similar fashion, it can be shown that the average propagator
for random rotations acting on the $8\IZ^1\IZ^2\IZ^3$ term is
\begin{equation}
\OL{e^{-2\UPSS (\chi^1 \IZ^1 + \chi^2 \IZ^2 + \chi^3 \IZ^3)}} ~=~
e^{-\frac{t}{2} (c^{11} + c^{22} + c^{33} + 8 c^{12} \IZ^1 \IZ^2
+ 8 c^{13} \IZ^1 \IZ^3 + 8 c^{23} \IZ^2 \IZ^3)}
\end{equation}
These averages are easily rotated back
to the average of the original propagator
$\MAT{R}_{\delta^1\delta^2\delta^3}(\chi^1,\chi^2,\chi^3)$
simply by replacing $\IZ$ by $\IX$ throughout.
All the terms obtained by the above expansions
are collected in the next section.

\section{Decoding and Error Correction} \label{sec:dec}
In order to write the ensemble average of
Eq.\ (\ref{eq:product}) in a compact form,
define the time-dependent (nonunitary) operators
\begin{equation}
F^j ~=~ F^j(t) ~\equiv~ e^{-t\,c^{jj}/2}
\end{equation}
(acting by scalar multiplication), and
\begin{equation}
\MAT{F}_\LAB{C}^{jk} ~=~ \MAT{F}_\LAB{C}^{jk}(t)
~\equiv~ e^{-t\,c^{jk}4\IX^j\IX^k}
\end{equation}
(acting by left-multiplication).
Then the density matrix after decoherence becomes
\begin{equation} \label{eq:rhoC}
\RHO_\LAB{C} ~\equiv~
(\alpha + \beta 8 \IX^1 \IX^2 \IX^3)\,
{\cal F}_\LAB{C} \left[ \FRAC{1}{8} (1 + F^1 2\IZ^1)
(1 + F^2 2\IZ^2) (1 + F^3 2\IZ^3) \right]
(\tilde\alpha + \tilde\beta 8 \IX^1 \IX^2 \IX^3) ~,
\end{equation}
where ${\cal F}_\LAB{C}$ is a linear operator-valued function
defined on the products of $\IZ^k$ operators by
\begin{eqnarray}
{\cal F}_\LAB{C}[\,1\,] &\equiv& 1 ~,
\nonumber \\
{\cal F}_\LAB{C}[2\IZ^j] &\equiv& 2\IZ^j
\quad (1 \le j \le 3) ~,
\nonumber \\
{\cal F}_\LAB{C}[4\IZ^j\IZ^k] &\equiv&
\MAT{F}_\LAB{C}^{jk} \, 4\IZ^j\IZ^k
\quad (1 \le j < k \le 3)
\\ \nonumber
&\equiv& e^{-t\,c^{jk} 4\IX^j\IX^k} \, 4\IZ^j\IZ^k ~,
\\ \nonumber
{\cal F}_\LAB{C}[8\IZ^1\IZ^2\IZ^3] &\equiv& \MAT{F}_\LAB{C}^{12}
\MAT{F}_\LAB{C}^{13} \MAT{F}_\LAB{C}^{23} \, 8\IZ^1\IZ^2\IZ^3
\\ \nonumber
&\equiv& e^{-t\,c^{12} 4\IX^1\IX^2 - t\,c^{13} 4\IX^1\IX^3
- t\,c^{23} 4\IX^2\IX^3} \, 8\IZ^1\IZ^2\IZ^3 ~.
\end{eqnarray}

Next, the decoding operation is performed by applying the
same two c-NOT's used in encoding, i.e. $\MAT{S}^{23|1}$.
On inserting appropriate factors of unity,
the resulting density matrix can be written as
\begin{eqnarray} \label{eq:fd_def}
\RHO_\LAB{D} &\equiv& ( \alpha + \beta 2\IX^1 )
{\cal F}_\LAB{D}\left[ \FRAC{1}{8}\, \MAT{S}^{23|1} (1 + F^1 2\IZ^1)
(1 + F^2 2\IZ^2) (1 + F^3 2\IZ^3)\, \MAT{S}^{23|1}
\right] ( \tilde\alpha + \tilde\beta 2\IX^1 ) \quad \\
&=& ( \alpha + \beta 2\IX^1 )
{\cal F}_\LAB{D}\left[ \FRAC{1}{8}\, (1 + F^1 2\IZ^1)
(1 + F^2 4\IZ^1\IZ^2) (1 + F^3 4\IZ^1\IZ^3) \right]
( \tilde\alpha + \tilde\beta 2\IX^1 ) ~, \nonumber
\end{eqnarray}
where the linear operator-valued function
${\cal F}_\LAB{D}$ is is defined on
the products of $\IZ^k$ operators by
\begin{equation}
{\cal F}_\LAB{D}[\, \MAT{S}^{23|1} \, \MAT{X} \, \MAT{S}^{23|1} \,] ~=~
\MAT{S}^{23|1} \, {\cal F}_\LAB{C}[\, \MAT{X} \,] \, \MAT{S}^{23|1} ~.
\end{equation}
This translates to
\begin{eqnarray} \label{eq:fcofd}
{\cal F}_\LAB{D}[\,1\,] &\equiv& 1 ~,
\nonumber \\
{\cal F}_\LAB{D}[2\IZ^1] &\equiv& 2\IZ^1 ~,
\nonumber \\
{\cal F}_\LAB{D}[2\IZ^2] &\equiv& \MAT{F}_\LAB{D}^{12} \, 2\IZ^2
~\equiv~ e^{-tc^{12} 4\IX^1\IX^3} \, 2\IZ^2  ~,
\nonumber \\
{\cal F}_\LAB{D}[2\IZ^3] &\equiv& \MAT{F}_\LAB{D}^{13} \, 2\IZ^3
~\equiv~ e^{-tc^{13} 4\IX^1\IX^2} \, 2\IZ^3  ~,
\nonumber \\
{\cal F}_\LAB{D}[4\IZ^1\IZ^2] &\equiv& 4\IZ^1\IZ^2 ~,
\\ \nonumber
{\cal F}_\LAB{D}[4\IZ^1\IZ^3] &\equiv& 4\IZ^1\IZ^3 ~,
\\ \nonumber
{\cal F}_\LAB{D}[4\IZ^2\IZ^3] &\equiv& \MAT{F}_\LAB{D}^{23} \, 4\IZ^2\IZ^3
~\equiv~ e^{-tc^{23} 4\IX^2\IX^3} \, 4\IZ^2\IZ^3  ~,
\\ \nonumber
{\cal F}_\LAB{D}[8\IZ^1\IZ^2\IZ^3] &\equiv& \MAT{F}_\LAB{D}^{12}
\MAT{F}_\LAB{D}^{13} \MAT{F}_\LAB{D}^{23} \, 8\IZ^1\IZ^2\IZ^3
\\ \nonumber
&\equiv& e^{-tc^{12} 4\IX^1\IX^3 - tc^{13} 4\IX^1\IX^2 - tc^{23} 4\IX^2\IX^3}
\, 8\IZ^1\IZ^2\IZ^3  ~,
\end{eqnarray}
since
\begin{eqnarray} \label{eq:fsubd}
\MAT{F}_\LAB{D}^{jk} &\equiv&
\MAT{S}^{23|1} e^{-t\,c^{jk}4\IX^j\IX^k} \MAT{S}^{23|1}
~=~ e^{-t\,c^{jk}\MAT{S}^{23|1} 4\IX^j\IX^k \MAT{S}^{23|1}}
\\ \nonumber &=&
e^{-t\,c^{jk} 4\IX^\ell\IX^m}
~=~ \left\{ \begin{array}{rl}
e^{-t\,c^{12} 4\IX^1\IX^3} & (j = 1, k = 2); \\
e^{-t\,c^{13} 4\IX^1\IX^2} & (j = 1, k = 3); \\
e^{-t\,c^{23} 4\IX^2\IX^3} & (j = 2, k = 3).
\end{array} \right.
\end{eqnarray}

Finally, the error correcting Toffoli gate $\MAT T^{1|23}
\equiv 2\IX^1\EM^2\EM^3 + (1 - \EM^2\EM^3)$ is applied.
Since $\MAT T^{1|23}$ commutes with the outer factors
$\alpha + \beta\, 2\IX^1$ and its conjugate,
one could define a new function and work out
how it operates on the transformed states
as was done with ${\cal F}_\LAB{D}$ above.
Rather than simply permuting product operators
as $\MAT S^{23|1}$ does, however, $\MAT T^{1|23}$
maps each to a linear combination of four products,
and in order to evaluate the partial trace over the
ancillae it is necessary to fully expand the result.
Fortunately, this laborious task can be avoided by noting
that for {\em any\/} product operator expression $\MAT X$,
the partial trace over the ancillae $4 \langle \MAT X \rangle^{23}$
is the same as $4 \langle {\cal E}[\MAT X] \rangle^{23}$
\begin{equation} \label{eq:proj}
\equiv~ 4 \left\langle \EP^2\EP^3 \,\MAT X\, \EP^2\EP^3 +
\EP^2\EM^3 \,\MAT X\, \EP^2\EM^3 + \EM^2\EP^3 \,\MAT X\, \EM^2\EP^3
+ \EM^2\EM^3 \,\MAT X\, \EM^2\EM^3 \right\rangle^{23}
\end{equation}
(This follows, for example, from the product
operator characterization of the partial trace
given in Ref.\ \cite{SomCorHav:98}, Eq.\ (40).)
In addition, the Toffoli commutes with
all of the idempotents $\EPM^j$ ($j = 2,3)$,
so that $\cal E$ can be applied
{\em before\/} applying the Toffoli gate.
As shall be seen momentarily, this
dramatically simplies the intermediate results,
so that the linear combinations created on applying
the final Toffoli are easily dealt with.

Because the idempotents $\EPM^j$ commute with
$\IZ^k$ for all $1 \le k \le j \le 3$, we can
apply $\cal E$ directly to the coefficients
$\MAT F_\LAB{D}^{jk}$ in Eq. (\ref{eq:fcofd}), e.g.
\begin{eqnarray}
{\cal E}[\MAT F_\LAB{D}^{12}] &=&
\sum_{\epsilon^2,\,\epsilon^3\in\{\pm1\}}
\MAT E_{\epsilon^2}^2 \MAT E_{\epsilon^3}^3
\left( \cosh(tc^{12}) - 4\IX^1\IX^3 \sinh(tc^{12}) \right)
\MAT E_{\epsilon^2}^2 \MAT E_{\epsilon^3}^3
\\ \nonumber &=&
\sum_{\epsilon^2,\,\epsilon^3\in\{\pm1\}} \left(
\cosh(tc^{12}) \left( \MAT E_{\epsilon^2}^2 \MAT
E_{\epsilon^3}^3 \right)^2 - 4\IX^1\IX^3 \sinh(tc^{12})
\left( \MAT E_{\epsilon^2}^2 \MAT E_{-\epsilon^3}^3 \right)
\left( \MAT E_{\epsilon^2}^2 \MAT E_{\epsilon^3}^3 \right)
\right)
\\ \nonumber &=&
\cosh(tc^{12}) \sum_{\epsilon^2,\,\epsilon^3\in\{\pm1\}}
\MAT E_{\epsilon^2}^2 \MAT E_{\epsilon^3}^3
~=~ \cosh(tc^{12}) ~\equiv~ F^{12} ~.
\end{eqnarray}
In a similar fashion, it can be shown that
\begin{eqnarray}
F^{13} ~\equiv~ {\cal E}[\MAT F_\LAB{D}^{13}] ~=~ \cosh(tc^{13}) ~,
&\qquad&
F^{23} ~\equiv~ {\cal E}[\MAT F_\LAB{D}^{23}] ~=~ \cosh(tc^{23}) ~,
\end{eqnarray}
and $F^{123} \equiv$
\begin{equation}
{\cal E}[\MAT F_\LAB{D}^{12}\MAT F_\LAB{D}^{13}\MAT F_\LAB{D}^{23}]
~=~ \cosh(tc^{12}) \cosh(tc^{13}) \cosh(tc^{23}) -
\sinh(tc^{12}) \sinh(tc^{13}) \sinh(tc^{23}) ~.
\end{equation}

Thus applying the Toffoli to the projection
of each term in Eq.\ (\ref{eq:fcofd}) yields
\begin{eqnarray} \label{eq:fcofe}
\MAT T^{1|23} {\cal E}[{\cal F}_\LAB{D}[\,1\,]]
\MAT T^{1|23} &=& 1 ~,
\nonumber \\
\MAT T^{1|23} {\cal E}[{\cal F}_\LAB{D}[2\IZ^1]] \MAT T^{1|23}
&=& \IZ^1 + 2\IZ^1\IZ^2 + 2\IZ^1\IZ^3 - 4\IZ^1\IZ^2\IZ^3 ~,
\nonumber \\
\MAT T^{1|23} {\cal E}[{\cal F}_\LAB{D}[2\IZ^2]] \MAT T^{1|23}
&=& F^{12} \, 2\IZ^2 ~,
\nonumber \\
\MAT T^{1|23} {\cal E}[{\cal F}_\LAB{D}[2\IZ^3]] \MAT T^{1|23}
&=& F^{13} \, 2\IZ^3 ~,
\\ \nonumber
\MAT T^{1|23} {\cal E}[{\cal F}_\LAB{D}[4\IZ^1\IZ^2]] \MAT T^{1|23}
&=& \IZ^1 + 2\IZ^1\IZ^2 - 2\IZ^1\IZ^3 + 4\IZ^1\IZ^2\IZ^3 ~,
\\ \nonumber
\MAT T^{1|23} {\cal E}[{\cal F}_\LAB{D}[4\IZ^1\IZ^3]] \MAT T^{1|23}
&=& \IZ^1 - 2\IZ^1\IZ^2 + 2\IZ^1\IZ^3 + 4\IZ^1\IZ^2\IZ^3 ~,
\\ \nonumber
\MAT T^{1|23} {\cal E}[{\cal F}_\LAB{D}[4\IZ^2\IZ^3]] \MAT T^{1|23}
&=& F^{23} \, 4\IZ^2\IZ^3 ~,
\\ \nonumber
\MAT T^{1|23} {\cal E}[{\cal F}_\LAB{D}[8\IZ^1\IZ^2\IZ^3]] \MAT T^{1|23} &=&
F^{123} \left( -\IZ^1 + 2\IZ^1\IZ^2 + 2\IZ^1\IZ^3 + 4\IZ^1\IZ^2\IZ^3 \right)
\end{eqnarray}
The partial trace over the ancilla simply drops all terms
from the above that contain factors of $\IZ^2$ or $\IZ^3$,
which eliminates all lines from the above equation containing
a doubly-indexed time-dependent coefficient $F^{jk}$.
Multiplying each of the surviving terms by the appropriate scalar
exponential $F^j \equiv \exp(-tc^{jj}/2)$ and combining finally yields
\begin{eqnarray} \label{eq:final_result}
\RHO_\LAB{E}^1 &\equiv& \left\langle \RHO_\LAB{E} \right\rangle^{23}
~\equiv~ \left\langle (\alpha + \beta\, 2\IX^1) \MAT T^{1|23}
\RHO_\LAB{D} \MAT T^{1|23} (\tilde\alpha + \tilde\beta\, 2\IX^1) \right\rangle^{23}
\nonumber \\
&\equiv& (\alpha + \beta\, 2\IX^1) \left\langle
\tfrac18 \, \MAT T^{1|23} {\cal E}[\, {\cal F}_\LAB{D} [\,
1 + F^1\, 2\IZ^1 + F^2\, 4\IZ^1\IZ^2 + F^3\, 4\IZ^1\IZ^3
\rule[0pt]{0pt}{14pt} \right. \\ \nonumber
&& \quad \left. \rule[0pt]{0pt}{14pt}
+\, F^1F^2F^3 \, F^{123} \, 8\IZ^1\IZ^2\IZ^3 \,] \,]
\MAT{T}^{1|23} \right\rangle^{23} (\tilde\alpha + \tilde\beta\, 2\IX^1)
\\ \nonumber
&=& (\alpha + \beta\, 2\IX^1) \,\HALF \left(
1 + (F^1 + F^2 + F^3 - F^1F^2F^3 \, F^{123}) \IZ^1
\right) (\tilde\alpha + \tilde\beta\, 2\IX^1)
\\ \nonumber
&=& \HALF + \Re(\tilde\alpha\beta) 2\IX^1 + \left( \Im(\tilde\alpha\beta)
2\IY^1 + (|\alpha|^2 - |\beta|^2) \IZ^1 \right) \Theta(t) ~,
\end{eqnarray}
where
\begin{eqnarray}
\Theta(t) &\equiv& \HALF\, (\, F^1 + F^2 + F^3 - F^1F^2F^3 \, F^{123} \,)
\nonumber \\
&\equiv& \HALF\, (\, e^{-tc^{11}/2} + e^{-tc^{22}/2} + e^{-tc^{33}/2} \,)
- \tfrac18 \, e^{-tc^{11}/2} \, e^{-tc^{22}/2} \, e^{-tc^{33}/2} \, \times
\\ \nonumber
&& \quad\left( e^{-tc^{12} - tc^{13} - tc^{23}} +
e^{tc^{12} + tc^{13} - tc^{23}} + e^{tc^{12} - tc^{13} + tc^{23}}
+ e^{-tc^{12} + tc^{13} + tc^{23}} \right) ~.
\end{eqnarray}
This equation describes how the error-corrected
state $\RHO_\LAB{E}^1$ of the data spin \#$1$
decays in the presence of decoherence due to
external random fields about the $\LAB{x}$-axis,
where $t\,c^{jj}$, $t\,c^{jk}$ are the variances and
covariances among the phases of the three spins, as described
in section {\bf \ref{sec:avg}} (Eq.\ (\ref{eq:cov_def})).

It is readily verified that the derivative $\dot\Theta(t)$
vanishes at $t = 0$ regardless of the covariances, as expected
from the analysis given in section {\bf \ref{sec:qec}}.
Two special cases are of particular interest in the following.
The first occurs when the random phases $\chi^k$
are identically distributed and uncorrelated,
so that $c^{11} = c^{22} = c^{33} \equiv 2/\tau$
and $c^{12} = c^{13} = c^{23} = 0$;
then the above simplifies to:
\begin{equation} \label{eq:uncorr_result}
\Theta(t) ~=~ \HALF\, \left(\, 3 e^{-t/\tau} - e^{-3t/\tau} \,\right)
\end{equation}
The second occurs when these random variables are totally correlated,
so that $c^{jk} \equiv 2/\tau$ for $1 \le j \le k \le 3$; then:
\begin{equation} \label{eq:correl_result}
\Theta(t) ~=~ \tfrac18\, \left(\, 9 e^{-t/\tau} - e^{-9t/\tau} \,\right)
\end{equation}

\section{Discussion of Theoretical Results} \label{sec:dis}
Quantum error correcting codes were designed to
prevent decoherence from occuring within a subsystem, 
at the expense of increasing decoherence in a quantum
environment with which it interacts in a controllable fashion.
In order to devise a quantum error correcting code,
the mechanism of decoherence must be precisely known.
Since a code can only correct for decay due to decoherence to
first order (in the absence of concatenation \cite{Preskill:99}),
decoherence can only be prevented over long time
intervals if the error correction procedure can be
performed much more rapidly than decoherence occurs.
To do this, it is necessary to be able to
rapidly reset the ancillae to their ground states
after each correction so they can be reused,
or else to have a large supply of ancillae in their
ground states that can rapidly replace the used ones.
Neither of these conditions is true in liquid-state
NMR, since $T_1 \ge T_2$ and the number of spins whose
interactions can be precisely controlled is limited.

Of course, pure states are not available
at all in liquid-state NMR spectroscopy,
as was assumed above for the ancillae.
The isomorphism which exists between the
dynamics of pure and pseudo-pure states
ensures that error correction will nevertheless work,
so long as the ancillae are in a pseudo-pure
ground state relative to the data spin.
As discussed in detail elsewhere \cite{HaSoTsCo:99,Warren:97},
the preparation of the ancillae in a pseudo-pure state
entails a loss of polarization that exceeds any potential gain
through error correction, so that the net signal-to-noise in the
NMR spectrum is actually lowered by the use of error correction.
It would therefore be advantageous if a broader class of mixed
states for the ancillae were also suitable for error correction,
but arguments are given in an Appendix which imply that this
is probably not true, and certainly not for any diagonal
($\LAB{z}$-polarized) mixed state of the ancillae.

In NMR, where decoherence occurs via a Raman process
involving both translational and rotational molecular motions,
measurements of the covariances in the fluctuating fields
are important indicators of the nature of these motions.
Thus, even though error correction cannot be used to sharpen
the lines or improve signal-to-noise in NMR spectroscopy,
the idea of coherently mixing together single and multiple
quantum coherences so as to cancel the first-order decay,
under certain {\em specific\/} assumptions regarding the
underlying mechanisms responsible for the relaxation
and the correlations among them, clearly has the potential
of being useful in NMR studies of molecular motions.
The higher moments of the initial decay curve should
also provide further confirmation of the assumptions.
For example, the second derivative $\ddot\Theta(t)$
at $t=0$ is $\ddot\Theta(0)$
\begin{equation}
=~ -\tfrac14 \left( 2(c^{12})^2 + 2(c^{13})^2 + 2(c^{23})^2
+ c^{11}c^{22} + c^{11}c^{33} + c^{22}c^{33} \right) ~<~ 0 ~,
\end{equation}
while the third derivative is $\dot{\ddot\Theta}(0)$
\begin{equation} \begin{split}
=~ & \tfrac1{16} \left( 3(c^{11})^2(c^{22}+c^{33}) + 3(c^{22})^2(c^{11}+c^{33})
+ 3(c^{33})^2(c^{22}+c^{33}) + 6(c^{11}c^{22}c^{33})
\right. \\ & +\, \left.
12((c^{12})^2+(c^{13})^2+(c^{23})^2)(c^{11}+c^{22}+c^{33})
+ 48(c^{12}c^{13}c^{23})
\right) ~,
\end{split} \end{equation}
which we expect will be generally positive.

Decoherence is of central importance to quantum mechanics,
since it is the process by which classical statistical
mechanics emerges from the underlying deterministic
evolution of wave functions \cite{GiuliniEtAl:96,Zurek:91}.
It is also a process which has proven extremely challenging
to study in its full generality \cite{Zurek:98,Haroche:98}.
Although again somewhat limited by its ensemble nature,
NMR is both theoretically and experimentally a very
convenient model system in which to study decoherence.
Because quantum codes only protect against decoherence
by specific mechanisms, they can be used to design NMR
experiments to test the validity of theoretical models.
This will now be illustrated by describing liquid-state NMR
experiments which validate the forgoing theoretical results.

\section{Gradient Implementation of Decoherence} \label{sec:grd}
Loss of phase coherence is, of course,
a natural relaxation process in NMR spectroscopy.
The longitudinal fluctuating fields that induce
this decoherence, however, come from many different
sources, both intramolecular and intermolecular,
and involving both spin and non-spin degrees of freedom.
Superimposed on these longitudinal fluctuations
are transverse fluctuations, which involve
an exchange of energy with the environment.
The three-bit code majority logic analyzed above can
deal with fluctuations about only one transverse axis;
by including $\pi/2$ rotation and its inverse at the end
of the encoding and beginning of the decoding, respectively,
it can deal instead with longitudinal fluctuations.
(Fluctuations about all three axes can
only be corrected using at least four ancillae
\cite{KnillLafla:97,Gottesman:98,Steane:98b,Preskill:99}.)
In addition, intramolecular dipole-dipole interactions
are an important source of decoherence in liquid-state NMR,
and these will also not be corrected by the three-bit code.
Thus the forgoing theoretical results can only be
rigorously demonstrated by experiments in which the
dominant relaxation mechanism is a known, artificially
induced fluctuation about a single transverse axis.

This is done here using molecular diffusion
to randomize the spatial variations in the
phase created by a pulsed field gradient.
A field gradient along the $\LAB{z}$-axis causes
the spins to precess at different rates,
depending on their z-coordinates.
This winds the transverse magnetization
due to each coherence into a spiral,
whose pitch decreases linearly with
the length of the gradient pulse.
In a liquid sample, the decay of a tightly wound spiral
is due almost entirely to diffusion along its axis.
Because the change in phase due to diffusion
is exactly the same for every spin in a molecule,
this implements the totally correlated error model.

The quantitative analysis of this process is best done
using the $k$-space formalism \cite{SodickCory:98}.
The Fourier transform of the distribution of
transverse magnetization $\upsilon(z)$ along the
$\LAB{z}$-axis will be denoted here by $\Upsilon(k)$.
The spiral produced by the gradient is described by
the product $\upsilon(z) \exp( \imath n_{ij} k_0 z )$,
where $n_{ij}$ is the order of the coherence in question,
whose Fourier transform is $\Upsilon(k) \delta( n_{ij} (k - k_0) )$.
Diffusion operates on the magnetization distribution by
convolution with a Gaussian whose variance is $\sigma^2 = D t$,
where $D$ is the diffusion coefficient \cite{SodickCory:98},
i.e.\ $\upsilon(z) \exp( \imath n_{ij} k_0 z ) \star
\exp( -z^2 / (2\sigma) ) / \sqrt{2\pi\sigma^2}$.
The Fourier transform is thus multiplied by the
Gaussian $\exp( -(k - k_0)^2 n_{ij}^2 D / 2 )$,
so that an inverse gradient pulse produces an echo
which decays with the square of the coherence order,
as expected from a totally correlated dephasing process.
These relations may be summarized as follows:
\begin{equation*} \begin{picture}(390,120)(10,0)
\stepcounter{equation} \put(430,55){(\theequation)}
\put(0,90){$\upsilon(z)$} \put(36,96){\sf gradient}
\put(30,90){\vector(1,0){55}} \put(91,90){$\upsilon(z)
e^{ -\imath n_{ij} k_0 z }$} \put(187,90){\vector(1,0){55}}
\put(192,96){\sf diffusion} \put(248,90){$\upsilon(z) e^{ -\imath %
n_{ij} k_0 z } \star e^{ -z^2 / 2 \sigma^2 } / \sqrt{2\pi\sigma^2}$}
\put(12,48){\vector(0,-1){24}} \put(6,50){\sf FT}
\put(12,60){\vector(0,1){24}}
\put(103,48){\vector(0,-1){24}} \put(97,50){\sf FT}
\put(103,60){\vector(0,1){24}}
\put(260,48){\vector(0,-1){24}} \put(254,50){\sf FT}
\put(260,60){\vector(0,1){24}}
\put(0,10){$\Upsilon(k)$} \put(36,16){\sf gradient}
\put(30,10){\vector(1,0){55}} \put(91,10){$\Upsilon(k)
\delta( n_{ij} (k - k_0) )$} \put(187,10){\vector(1,0){55}}
\put(192,16){\sf diffusion} \put(254,10){$\Upsilon(k) \star%
\delta( n_{ij} (k - k_0) ) e^{ -c_{ij}^2 k^2 \sigma^2 / 2 }$}
\end{picture} \end{equation*}

Implementation of the uncorrelated error model was done
by applying three successive gradient-diffusion sequences.
In each sequence, two of the spins were refocussed
during a gradient-pulse by selective RF $\pi$-pulses,
while maintaining the phase ramp on the third spin
(which was different in each of the three sequences).
This was followed by a time interval to allow
diffusion to randomize the phase of the third spin,
after which a second gradient and further selective
$\pi$-pulses were used to refocus all three spins
(save for the coherence of the third lost to diffusion).
This procedure randomized the phase of each spin equally
and independently during separate diffusion intervals,
as required by the uncorrelated model.
The complex RF pulse and gradient sequences which
performed this task are described in the following section.

\section{Experimental Implementation of Error Correction} \label{sec:exp}
This section provides a detailed description of the
RF and gradient pulse sequences used to implement
the three-bit quantum error correcting code
(see Fig.\ \ref{fig:diagram}) as well as the
correlated and uncorrelated decoherence models.
The challenge here lies in obtaining precisely
the desired effective Hamiltonian at each step,
and in putting these steps together without side effects.
In particular, the finite duration of all experimentally
realizable RF and gradient pulses allow the system to evolve
and so pick up wanted changes in the relative phases of its states.
Therefore these unwanted evolutions must be refocussed,
at the cost of increased complexity in the implementations.
Additionally, the system is subjected to incoherent
errors due to pulse imperfections and field inhomogeneity.
These could be compensated for by phase cycling and other
averaging techniques \cite{ErnBodWok:87,Freeman:98},
but this is undesirable in the present context since
it complicates the interpretation of the experiments.
Of course, such difficulties are not unique to NMR, but are
expected to varying degrees in any quantum information processor.
The intrinsically long decoherence times, innate averaging
over ensembles and superb coherent control available in modern
NMR spectroscopy is what makes these experiments possible today.

The three-bit quantum error correcting code was
realized with a sample of ${}^{13}\LAB{C}$-labeled alanine
($\LAB{NH}_3^+-\LAB{C}^\alpha\LAB{H}(\LAB{C}^\beta\LAB{H}_3)-\LAB{CO}_2^-$)
in $\LAB{D}_2\LAB{O}$ solution at room temperature.
The measurements were carried out on a Bruker AMX400
spectrometer ($9.6$ T) equipped with a $5$ mm probe
tuned to the ${}^{13}\LAB{C}$ and ${}^1\LAB{H}$
frequencies of $100.61$ and $400.13$ MHz, respectively.
This probe was equipped with a $\LAB{z}$-gradient coils
capable of generating field gradients of $60$ G / cm.
With decoupling of alanine's protons, this system
exhibits a weakly-coupled carbon NMR spectrum.
The internal Hamiltonian in the rotating frame is
\begin{equation}
\MAT H_\LAB{int} ~=~ \omega^\LAB{C'} \IZ^\LAB{C'}
+ \omega^\LAB{C^\alpha} \IZ^\LAB{C^\alpha}
+ \omega^\LAB{C^\beta} \IZ^\LAB{C^\beta}
+ 2\pi J^\LAB{C'C^\alpha} \IZ^\LAB{C'}\IZ^\LAB{C^\alpha}
+ 2\pi J^\LAB{C'C^\beta} \IZ^\LAB{C'}\IZ^\LAB{C^\beta}
+ 2\pi J^\LAB{C^\alpha C^\beta} \IZ^\LAB{C^\alpha}\IZ^\LAB{C^\beta}
\end{equation}
where (with the transmitter on-resonance with the $\LAB{C}^\alpha$)
\begin{eqnarray}
\omega^\LAB{C'} / (2\pi) ~=~ 12580 \mathrm{Hz}, & 
\omega^\LAB{C^\alpha} / (2\pi) ~=~ 0 \mathrm{Hz}, & 
\omega^\LAB{C^\beta} / (2\pi) ~=~ -3443 \mathrm{Hz}, \\
J^\LAB{C'C^\alpha} ~=~ 54.2 \mathrm{Hz}, &
J^\LAB{C'C^\beta} ~=~ 1.2 \mathrm{Hz}, &
J^\LAB{C^\alpha C^\beta} ~=~ 35.1 \mathrm{Hz} ~. \nonumber
\end{eqnarray}
The $\LAB{C^\alpha}$ was chosen as the data spin \#1,
and exhibits a well-resolved quartet in the spectrum.
This has the advantage that none of the gates
required for the error correction procedure
needed to use the small $1.2$ Hz coupling directly
(although it was necessary to refocus this coupling).
The carbonyl $\LAB{C'}$ (spin \#2) and $\LAB{C^\beta}$
(spin \#3) were used for the two ancilla spins.

To implement a universal set of quantum logic gates in
a spin system it is sufficient that all the spins be connected
by coupling pathways and sufficiently well-resolved to enable
arbitary rotations to be applied to any subset of the spins.
Here selective excitations were implemented using
phase-modulated Gaussian pulses \cite{Patt:92}.
These shaped pulses consisted of 384 complex
points with a total duration of $1.5$ msec,
with an excitation profile that consisted
of Gaussians centered on the frequencies
of the (one or more) spins of interest
with a standard deviation of $500$ Hz.
The transmitter was placed on the data spin,
enabling it to be controlled by soft rectangular pulses.

\begin{figure}[H] \begin{center}
\begin{picture}(450,300)
\put(-40,300){
\scalebox{0.85}[0.75]{\includegraphics[angle=270]{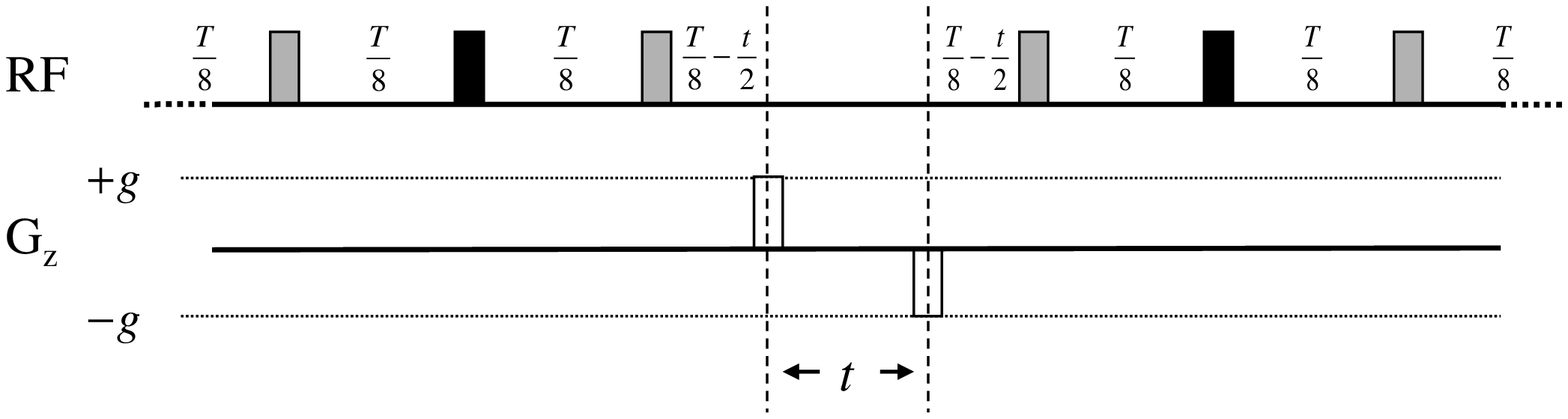}}
} \put(-45,150){
\scalebox{0.80}[0.80]{\includegraphics[angle=270]{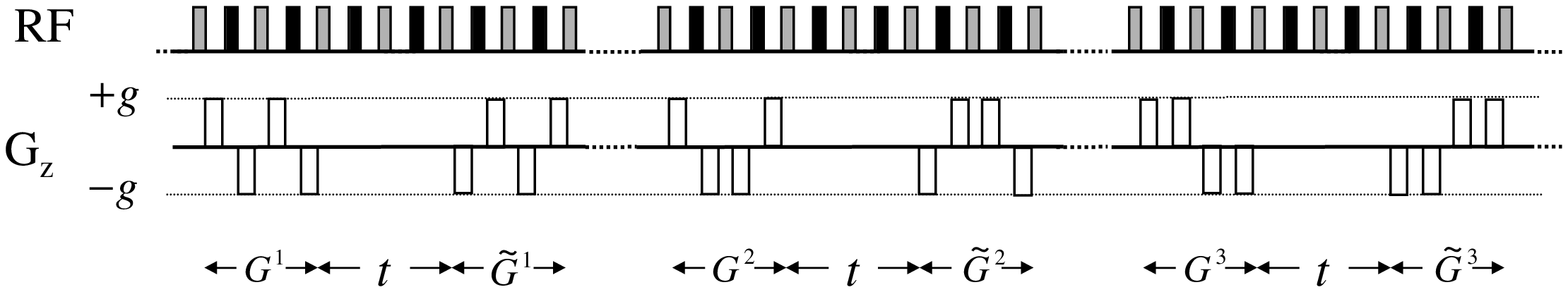}}
} \put(460,185){\textbf{(a)}} \put(460,40){\textbf{(b)}}
\end{picture}
\bigskip \figcap{
RF and gradient pulse sequence diagrams to implement both the
totally correlated (a) and uncorrelated (b) decoherence models.
In the figure, black boxes indicate $(\pi)^{12}$ RF pulses,
and grey boxes indicate $(\pi)^{13}$ pulses, while the
graph on the $G_z$ line indicates the gradient polarity.
The remaining symbols are defined in the main text.
} \label{fig:pulse_seq}
\end{center} \end{figure}

It is convenient to define five specific pulse sequences,
which were designed to yield a simple, well-defined
effective Hamiltonian and served as the ``modules''
(building-blocks) for the overall pulse sequences.
In the following list, all the time intervals are calculated
from the midpoints of their surrounding shaped pulses.
\begin{Desc}
\item[Identity$(t)$]
This is a ``time-suspension'' sequence which is designed
to refocus all components of the Hamiltonian by the
symmetric application of $\pi$-pulses \cite{ErnBodWok:87}.
Identity sequences correct the phase errors due
to the finite duration of RF and gradient pulses,
and are used whenever two noncommuting
operations must be applied sequentially.
The implementation used in these experiments is given by
the pulse sequence (in left-to-right temporal order)
\begin{equation}
\left(\pi\rule[2pt]{0pt}{14pt}\right)^{k \ell} -
\left(\pi\rule[2pt]{0pt}{14pt}\right)^{\ell m} -
\left(-\pi\rule[2pt]{0pt}{14pt}\right)^{k \ell} -
\left(\Delta\rule[2pt]{0pt}{14pt}\right) -
\left(\pi\rule[2pt]{0pt}{14pt}\right)^{k \ell} -
\left(-\pi\rule[2pt]{0pt}{14pt}\right)^{\ell m} -
\left(-\pi\rule[2pt]{0pt}{14pt}\right)^{k \ell} ~,
\end{equation}
where $(\pi)^{k\ell}$ denotes a $\pi$-rotation
of spins $k,\ell$ about the $\LAB{x}$-axis,
and similarly for $(\pi)^{\ell m}$ and $(\pi)^{k m}$.
These pulses have a duration of $\Delta$,
so that the net propagator can be written as e.g.
\begin{equation} \label{eq:eff_pul_pro}
\left(\pi\rule[2pt]{0pt}{14pt}\right)^{k \ell}
\quad\Leftrightarrow\quad
e^{-\UPSS\MAT H_\LAB{int}\Delta/2}
e^{-\UPSS\pi(\IX^k+\IX^\ell)}
e^{-\UPSS\MAT H_\LAB{int}\Delta/2}
\end{equation}
with similar expressions for the other pulses.
Thus if one places such an identity
sequence between two noncommuting pulses,
the evolution during the $\Delta/2$ duration of the last
half of the first pulse and the first half of the last
pulse (as in Eq.\ (\ref{eq:eff_pul_pro})) cancels with
the evolution during the intervening identity sequence.
Note that due to the symmetry of this sequence
all the $\LAB{CH}$ couplings are refocussed,
so that additional decoupling of the protons is not needed.
\item[Jdelay$(k,\ell,t)$]
This module yields the effective
Hamiltonian $2\pi J^{kl} \IZ^k\IZ^\ell$
where $k, l = 1,2,3$ are spin indices.
The parameter $t$ determines the net
acquired phase $\varphi = 2\pi J^{k\ell} t$;
if $k = 1$, $\ell = 2$, for example,
a phase shift of $\varphi$ is obtained from
$t = \varphi/(2J^{12}) = \varphi / (2 \times 54.2)$ sec.
The complete sequence is
\begin{eqnarray}
&& \left(\frac{t}4\right) \text{---}
\left(\pi\rule[2pt]{0pt}{14pt}\right)^{123} \text{---}
\left(\frac{t}4\right) \text{---}
\left(\pi\rule[2pt]{0pt}{14pt}\right)^{k\ell} \text{---}
\left(\frac{t}4\right) \text{---}
\left(-\pi\rule[2pt]{0pt}{14pt}\right)^{123}
\\ && \text{---} \, \nonumber
\left(\frac{t}4\right) \text{---}
\left(\Delta\rule[2pt]{0pt}{14pt}\right) \text{---}
\left(\pi\rule[2pt]{0pt}{14pt}\right)^{km} \text{---}
\left(-\pi\rule[2pt]{0pt}{14pt}\right)^{k\ell} \text{---}
\left(-\pi\rule[2pt]{0pt}{14pt}\right)^{km} ~,
\end{eqnarray}
where $[\Delta]$ is again a delay of duration
equal to that of the shaped pulses ($1.5$ msec.).
The finite duration of the pulses sets
a lower limit for the phase evolution
of $\varphi > 4\cdot2\cdot J^{k\ell}\Delta$
($= 0.21\pi$ for $J^{12}$ and $0.13\pi$ for $J^{13}$).
\item[JdelayInv$(k,\ell,t)$]
This is a short version of \textit{Jdelay} with one modification:
after its application all three spins are inverted
by $\pi$ relative to the result of \textit{Jdelay}.
Successive evolutions under $2\pi J^{12} \IZ^1\IZ^2$ and
$2\pi J^{13} \IZ^1\IZ^3$ were often needed in these experiments,
and the same result is obtained from using two successive
\textit{JdelayInv} modules as from two successive Jdelay modules.
The former, however, saves three shaped pulses, 
as may be seen from its pulse sequence:
\begin{equation}
\left(\frac{t}4\right) \text{---}
\left(\pi\rule[2pt]{0pt}{14pt}\right)^{k\ell} \text{---}
\left(\frac{t}4\right) \text{---}
\left(\pi\rule[2pt]{0pt}{14pt}\right)^{123} \text{---}
\left(\frac{t}4\right) \text{---}
\left(-\pi\rule[2pt]{0pt}{14pt}\right)^{k\ell} \text{---}
\left(\frac{t}4\right)
\end{equation}
\item[TC-Decohere$(g, \delta, t, T)$]
This module was used to induce decoherence
under totally correlated external random fields.
It includes two gradients of equal strength $g$
and duration $\delta$ but opposite polarity,
embedded in a time-suspension sequence of length $T$.
The gradients are separated by a period
$t$ during which diffusion takes place.
The effects of decoherence due to $T_2$ relaxation during
this module were reduced to a constant factor by keeping
the total time required fixed at $T$ while varying $t$.
Letting $[\LAB{z}\mathrm{-grad}(g,\delta)]$
denote a gradient pulse along the $\LAB{z}$-axis
of strength $g$ and duration $\delta$,
the pulse sequence used for this module was
\begin{eqnarray}
&& \left(\frac{T}8\right) \text{---}
\left(\pi\rule[2pt]{0pt}{14pt}\right)^{km} \text{---}
\left(\frac{T}8\right) \text{---}
\left(\pi\rule[2pt]{0pt}{14pt}\right)^{k\ell} \text{---}
\left(\frac{T}8\right) \text{---}
\left(-\pi\rule[2pt]{0pt}{14pt}\right)^{km} \text{---}
\left(\frac{T}8-\frac{t}2-\frac\delta2\right)
\\ && \text{---} \, \nonumber
\left(\LAB{z}\mathrm{-grad}(g,\delta)\rule[2pt]{0pt}{14pt}\right) \text{---}
\left(\frac{t}2-\frac\delta2\right) \text{---}
\left(\Delta\rule[2pt]{0pt}{14pt}\right) \text{---}
\left(\frac{t}2-\frac\delta2\right) \text{---}
\left(\LAB{z}\mathrm{-grad}(-g,\delta)\rule[2pt]{0pt}{14pt}\right)
\\ && \text{---} \, \nonumber
\left(\frac{T}8-\frac{t}2-\frac\delta2\right) \text{---}
\left(\pi\rule[2pt]{0pt}{14pt}\right)^{km} \text{---}
\left(\frac{T}8\right) \text{---}
\left(-\pi\rule[2pt]{0pt}{14pt}\right)^{k\ell} \text{---}
\left(\frac{T}8\right) \text{---}
\left(-\pi\rule[2pt]{0pt}{14pt}\right)^{km} \text{---}
\left(\frac{T}8\right)
\end{eqnarray}
This pulse sequence is shown in diagrammatic
form in Fig.\ \ref{fig:pulse_seq}(a).
\item[UC-Decohere$(g, \delta, r_2, r_1 )$]
This module was used to induce decoherence under
uncorrelated external random fields (see above).
It uses a combination of refocusing and gradient
pulses of absolute strength $g$ and duration $\delta$.
These gradient pulses were strung together into
one of three sequences, all of the form
\begin{eqnarray}
G^i: && \left(\LAB{z}\mathrm{-grad}(g_1^i,\delta)
\rule[2pt]{0pt}{14pt}\right) \text{---} \left(
\pi \rule[2pt]{0pt}{14pt} \right)^{12} \text{---}
\left(\LAB{z}\mathrm{-grad}(g_2^i,\delta)
\rule[2pt]{0pt}{14pt}\right) \text{---} \left(
\pi \rule[2pt]{0pt}{14pt} \right)^{13} \text{---}
\\ && \nonumber
\left(\LAB{z}\mathrm{-grad}(g_3^i,\delta)
\rule[2pt]{0pt}{14pt}\right) \text{---} \left(
\pi \rule[2pt]{0pt}{14pt} \right)^{12} \text{---}
\left(\LAB{z}\mathrm{-grad}(g_4^i,\delta)
\rule[2pt]{0pt}{14pt}\right) \text{---}
\left( \pi \rule[2pt]{0pt}{14pt} \right)^{13} ~,
\end{eqnarray}
where the polarities in the sequence $g^i$
for dephasing the $i$-th spin were given by
\begin{eqnarray}
g^1: && \left[ +g,\, -g,\, +g,\, -g \right] \nonumber \\
g^2: && \left[ +g,\, -g,\, -g,\, +g \right] \\ \nonumber
g^3: && \left[ +g,\, +g,\, -g,\, -g \right] ~.
\end{eqnarray}
Each of these gradient sequences, in turn, was
embedded in a fixed number of repetitions of
a sequence of refocusing pulses, of the form
\begin{equation} \label{eq:refocus}
\left( \delta \rule[2pt]{0pt}{14pt} \right) \text{---}
\left( \pi \rule[2pt]{0pt}{14pt} \right)^{12} \text{---}
\left( \delta \rule[2pt]{0pt}{14pt} \right) \text{---}
\left( \pi \rule[2pt]{0pt}{14pt} \right)^{13} ~,
\end{equation}
where $\left(\delta\right)$ indicates a time delay.
Note that two consecutive repetitions of this sequence
refocuses both the chemical shift and coupling evolution,
and that the sequences $G^i$ above with $g = 0$ comprise
two such repetitions and so do nothing, as desired.
A given number $R$ of repetitions of this pulse
sequence will be denoted by $( R )$,
while the gradient sequence which refocuses the
effects of the gradient sequence $( G^i )$ in a
given experiment will be denoted by $( \tilde{G}^i )$.
Depending on whether the intervening number of
refocusing sequences (\ref{eq:refocus}) is even
or odd, this will either be $( G^i )$ or the
same sequence with all its polarities reversed.
Thus the overall sequence used to independently
decohere all three spins and thereby implement
the uncorrelated decoherence model is given by
\begin{eqnarray} &&
\left( r_2 - r_1 \rule[2pt]{0pt}{14pt} \right) \text{---}
\left( G^1 \rule[2pt]{0pt}{14pt} \right) \text{---}
\left( r_1 \rule[2pt]{0pt}{14pt} \right) \text{---}
\left( \tilde{G}^1 \rule[2pt]{0pt}{14pt} \right) \text{---}
\left( r_2 - r_1 \rule[2pt]{0pt}{14pt} \right) \text{---}
\left( G^2 \rule[2pt]{0pt}{14pt} \right)
\\ \nonumber && \text{---}
\left( r_1 \rule[2pt]{0pt}{14pt} \right) \text{---}
\left( \tilde{G}^2 \rule[2pt]{0pt}{14pt} \right) \text{---}
\left( r_2 - r_1 \rule[2pt]{0pt}{14pt} \right) \text{---}
\left( G^3 \rule[2pt]{0pt}{14pt} \right) \text{---}
\left( r_1 \rule[2pt]{0pt}{14pt} \right) \text{---}
\left( \tilde{G}^3 \rule[2pt]{0pt}{14pt} \right) ~,
\end{eqnarray}
where $0 < r_1 < r_2$ are integers and
$( r_1 )$ or $( r_2 - r_1 )$ denote the enclosed
number of repetitions of (\ref{eq:refocus}).
This pulse sequence is shown in diagrammatic
form in Fig.\ \ref{fig:pulse_seq}(b).
Observe the total length of each experiment is
constant, which keeps the effect of intrinsic
$T_2$ relaxation to a constant overall factor.
The signal loss due to RF inhomogeneity
during the application of the long train
of refocusing pulses is reduced by a CPMG type
phase modulation of length eight \cite{Freeman:98}.
These phase changes are carried out
during a single scan of each experiment,
and no phase cycling is performed across scans.
\end{Desc}
In the \textsf{TC-Decohere} module, the gradient strength
and duration were $g = 35.7$ Gauss/cm and $\delta = 2.5$ msec.,
respectively, while the total time was kept fixed at $T = 64.5$
msec.\ with a diffusion time increment of $t = 4$ msec.
In the \textsf{UC-Decohere} module, the gradient strength
used was $|g| = 12.2$ Gauss/cm, with $\delta = 2.078$ msec.,
which yields an increment of $7.156$ msec.\ over $r_2 = 21$ time points.


An account of how to design pulse sequences,
composed of these modules, for any desired effective Hamiltonian
may be found in our earlier work \cite{SomCorHav:98,PrSoDuHaCo:99}.
For the sake of completeness, however, the basic ideas are
repeated for the case of the error-correcting Toffoli gate.
This gate may be written in exponential form and expanded
into a product of commuting propagators, as follows:
\begin{eqnarray}
\MAT{T}^{1|23} &\equiv& 2 \IX^1 \EM^2 \EM^3 + (1 - \EM^2 \EM^3) ~=~
e^{\UPSS \pi (1/2 - \IX^1 \EM^2 \EM^3) \EM^2 \EM^3} \\ &=~& \nonumber
e^{\UPS\pi/8}\,e^{-\UPS\pi/4\IX^1}\,e^{-\UPS\pi/4\IZ^2}\,e^{-\UPS\pi/4\IZ^2}
\,e^{\UPS\pi/2\IX^1\IZ^2}\,e^{\UPS\pi/2\IX^1\IZ^3}\,e^{\UPS\pi/2\IZ^2\IZ^3}
\,e^{-\UPS\pi\IX^1\IZ^2\IZ^3}
\end{eqnarray}
The propagators in this expression could be
further expanded into products of propagators
directly implementable using the above modules.
In these experiments, however, the Toffoli gate is
followed immediately by a partial trace over the ancillae,
i.e.\ by observing the data spin while decoupling the ancillae.
Hence all propagators that operate only on the ancillae
can be eliminated with no effect on the final result.
Since the net phase is also unobservable, this
leaves only the product of the propagators
\begin{equation}
e^{-\UPS\pi/4\IX^1} \, e^{\UPS\pi/2\IX^1\IZ^2}\,
e^{\UPS\pi/2\IX^1\IZ^3} \, e^{-\UPS\pi\IX^1\IZ^2\IZ^3} ~.
\end{equation}
This in turn can be implemented by the sequence of propagators
of the effective Hamiltonians (in reverse temporal order):
\begin{equation}
e^{\UPS\pi/4\IX^1}\,e^{\UPS\pi/2\IZ^1}\,e^{\UPS\pi\IZ^1\IZ^2}\,
e^{-\UPS\pi\IY^1}\,e^{\UPS\pi/2\IZ^1\IZ^3}\,e^{-\UPS\pi/2\IY^1}\,
e^{\UPS\pi\IZ^1\IZ^2}\,e^{\UPS\pi/2\IX^1}\,e^{\UPS7\pi/2\IZ^1\IZ^2}\,
e^{-\UPS\pi\IY^1}\,e^{\UPS\pi/2\IZ^1\IZ^3}\,e^{-\UPS\pi/2\IY^1}
\end{equation}
These effective Hamiltonians can all be obtained
by combining the above modules with RF pulses.

The state $\IZ^1\EP^2\EP^3$ was prepared first,
after which the other states $\IX^1\EP^2\EP^3$
and $\IY^1\EP^2\EP^3$ used in the experiments
were obtained by rotating spin \#1.
Starting with the three-spin equilibrium state $\IZ^1+\IZ^2+\IZ^3$,
a pair of c-NOT's was applied to the ancillae conditional on the data spin:
\begin{equation}
\MAT S^{2|1} \MAT S^{3|1} \left( \IZ^1 + \IZ^2 + \IZ^3 \right)
\MAT S^{3|1} \MAT S^{2|1} ~=~ \IZ^1(1 + 2\IZ^2 + 2\IZ^3)
\end{equation}
This was then subjected to the pulse sequence
(given in temporal order)
\begin{equation} \label{eq:pp_proj}
\left(\FRAC\pi2\right)^1 -
\left(\FRAC1{4J^{12}}\right) -
\left(\FRAC\pi2\right)_{\pi/4}^1 -
\left(\LAB{x}\mathrm{-grad}\right) -
\left(\FRAC\pi2\right)_{\pi/4}^1 -
\left(\FRAC1{4J^{13}}\right) -
\left(\FRAC\pi2\right)_{\pi/2}^1 -
\left(\LAB{y}\mathrm{-grad}\right) ~,
\end{equation}
where $[1/(2J^{1k})]$ is an evolution under the effective
Hamiltonian $2\pi J^{1k} \IZ^1\IZ^k$ a time $1/(2J^{1k})$
($k = 2, 3$), i.e.\ \textit{Jdelay}$(1, k, t)$,
$(\LAB{x}\mathrm{-grad})$ is a magnetic field
gradient of $\partial B_\LAB{z} / \partial x$,
and similarly for $(\LAB{y}\mathrm{-grad})$.
This yields $3 \IZ^1 \EP^2 \EP^3$.
Note that this method of preparation reduces
the signal of the data spin by $25$\%
(as opposed to the $50$\% expected from the
equilibrium populations \cite{Warren:97});
the transformation in Eq.\ (\ref{eq:pp_proj}) also
has the interesting property of being a projection,
in that applying it a second time to the
resulting state does not change the state.

The final operation required is to observe
the partial trace over the ancillae at the end.
Because only single quantum coherences of
the form $\IX^k$ and $\IY^k$ are observable,
to implement the partial trace during acquisition
it is only necessary to prevent antiphase
magnetization with respect to the ancillae from
evolving into observable single quantum terms.
This was achieved by interspersing the sampling
of the signal with $\pi$-pulses on the ancillae,
thereby repeatedly refocusing these terms.
This RF irradiation caused a fixed upfield shift of $\sim12$
Hz due to the Bloch-Seigert effect \cite{Slichter:90}.

\section{Experimental Results and Discussion} \label{sec:res}
The error correcting code shown in Fig.\ \ref{fig:diagram}
was applied to the $\IX^1\EP^2\EP^3$,
$\IY^1\EP^2\EP^3$ and $\IZ^1\EP^2\EP^3$
states.
One can regard this code as a modification of
the classical majority logic code for protecting
against bit flip errors about the $\LAB{x}$-axis.
It follows that the $\IX^1\EP^2\EP^3$ should not
be affected by the gradient induced decoherence,
whereas both $\IY^1\EP^2\EP^3$ and $\IZ^1\EP^2\EP^3$
will decay exponentially at the same
rate in the absence of error correction.
These exponential decay rates were measured by
applying the gradient diffusion procedure directly
to the $\IY^1$ and $\IZ^1$ states, respectively.


In Figs.\ \ref{fig:z_plot} and \ref{fig:y_plot},
the decay of the magnetization of the data spin
is plotted with and without the error correction
procedure starting from the $\IZ^1\EP^2\EP^3$
and $\IY^1\EP^2\EP^3$ states, respectively.
In both these experiments the initial slope
of the error corrected curve tends to zero
as the decoherence time $t \rightarrow 0$,
in accord with theoretical predictions.
This was quantified by using the rate
$1/\tau$ of the uncorrected decay, estimated
from a linear least-squares fit to its logarithm,
to predict the corrected decay curve from equation for totally
correlated decoherence $(9 \exp( -t/\tau ) - \exp( -9t/\tau ))/8$
(see Eq.\ (\ref{eq:correl_result})).
The correlation coefficient between the measured and
predicted amplitudes were $0.9845$ and $0.9704$ for the
$\IZ^1\EP^2\EP^3$ and $\IY^1\EP^2\EP^3$ experiments, respectively.
In the figures, the error-corrected data points have
been scaled so that their mean-square value is the
same as that of the corresponding predicted points;
no other free parameters were needed for these fits.


The scatter seen in the data points,
typically about $\pm1$\% of the peak
intensity after averaging of 16 repetitions,
can be attributed primarily to RF field inhomogeneity,
particularly during the long period of induced decoherence.
In Fig.\ \ref{fig:x_plot}, we also show the amplitudes of
the peak starting from the $\IX^1\EP^2\EP^3$ data spin state,
from which it is evident that it is not significantly
affected by the overall error correction procedure.
The apparent decay rate of $\approx 0.2$ $\mathrm{sec}^{-1}$
can likewise be attributed primarily to RF field inhomogeneity,
which produced a small transverse magnetization after encoding.

Figure \ref{fig:u_plot} shows the results of a similar
set of experiments on the $\IZ^1\EP^2\EP^3$ state,
with independent diffusion intervals for each of the three spins
to implement uncorrelated decoherence, as previously described.
As in the correlated experiments, the decay rate $1/\tau$
obtained from a least-squares fit to the logarithm
of the decay of the $\IX^1\EP^2\EP^3$ state due to
this decoherence procedure was used to calculate the
theoretical curve $(3\exp(-t/\tau) - \exp(-3t/\tau))/2$
(see Eq.\ (\ref{eq:uncorr_result})),
and scaled the corresponding error-corrected data
to have the same root-mean-square as the theoretical
curve sampled at the corresponding time points.
Although the experimental difficulties of juxtaposing
three independent gradient-decoherence sequences
without picking up unwanted phase shifts induces
appreciable nonrandom scatter in the data points,
the fit is once again in accord with theoretical predictions.
Also shown with a dashed line is the decay curve
expected for totally correlated decoherence,
showing that these experiments are indeed
capable of distinguishing these two cases.

Together, these results provide strong support not
only for the theory of quantum error correction,
but also for the ability of pseudo-pure states
to reproduce the dynamics of true pure states.
This ability is important because liquid-state
NMR spectroscopy provides a degree of coherent
control that is presently more difficult to obtain
in other quantum systems of comparable complexity.
Although limited to about ten spins \cite{HaSoTsCo:99,Warren:97}
and hence within reach of classical computer simulations,
such an experimentally accessible paradigm for quantum
information processing should be quite useful particularly
in the study of decoherence \cite{GiuliniEtAl:96}.
It forces one to consider, and enables one
to experimentally determine, actual relaxation
superoperators, rather than working from
idealized theoretical models of decoherence.
The development of error correcting codes that can
handle such real-life decoherence is a significant
challenge whose solution is likely to be applicable
to quantum information processing in other systems.

Conversely, the theory of quantum error correction promises
to lead to new methods for studying molecular dynamics
through the relaxation of multiple quantum coherences
\cite{Redfield:65,WerbeGrant:77,VoldVold:78,%
WokauErnst:78,PengWagner:94,KumarKumar:96}.
This may be seen, for example, by considering the
second derivative of the error-corrected curves at
$t = 0$ in the uncorrelated and totally correlated
cases, which are $-3/\tau^2$ and $-9/\tau^2$,
respectively, while the corresponding inflection
points occur at $\ln(3)\tau/2$ and $\ln(3)\tau/4$.
Moreover, since the covariances of the random
fields at the three spins occur in our formulae
for the decay of the error-corrected states,
it would appear possible to derive these covariances
directly from sufficiently detailed measurements.
One of the goals of on-going studies of quantum
error correction by the present authors is to develop
codes which can more clearly reveal the nature
of the underlying spin and molecular dynamics,
for example by correcting for dipole-dipole relaxation.
Work in these directions is currently in progress.


\bigskip\centerline{\bf Acknowledgements}\smallskip
This work was supported by the U.\ S.\ Army Research
Office under grant number DAAG 55-97-1-0342 from
the DARPA Microsystems Technology Office.

\section*{Appendix: Error Correction in Mixed States}
\begin{figure}[H] \begin{center}
\begin{picture}(400,120)
\put(-45,128){ \includegraphics[scale=0.7,angle=270]{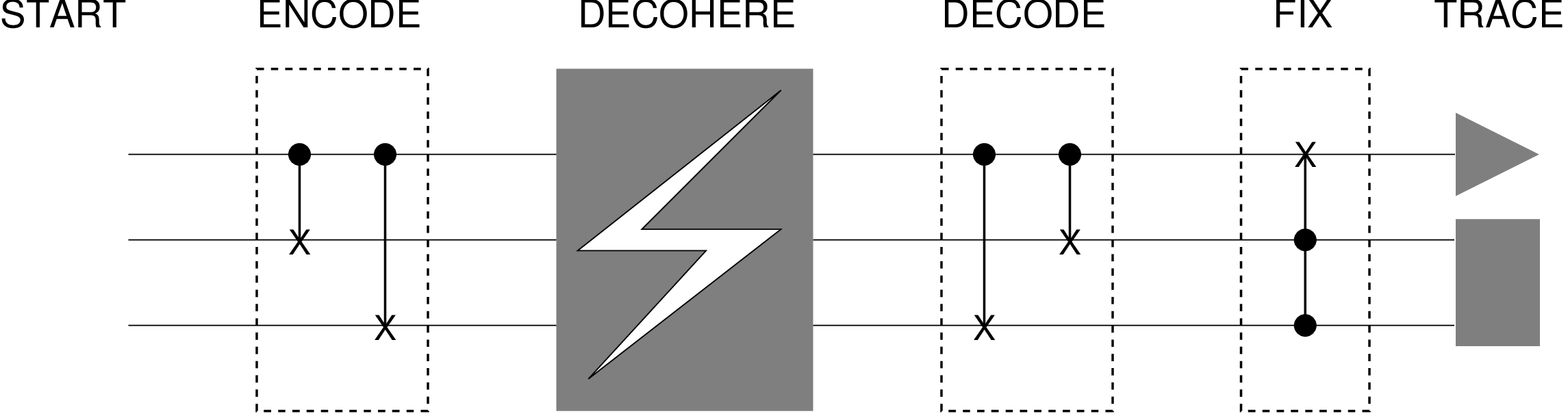}
} \put(-20,72){$\RHO_\LAB{A}^1$}
\put(-20,47){$\EP^2$} \put(-20,22){$\EP^3$}
\end{picture}
\bigskip \figcap{
Diagram giving an overview of the error correction procedure.
The three spins correspond to the three horizontal lines.
Vertical lines connecting them are quantum gates
(unitary operations), where ``$\LAB{x}$'' indicates
a target spin and ``{\Large$\bullet$}'' a control spin.
The first dashed box is the encoding step,
which consists of two successive c-NOT's (see text).
The filled box with the lightning bolt indicates decoherence
through random rotations about the $\LAB{x}$-axis.
The second dashed box is the decoding step,
while the third is the Toffoli gate which
corrects for decoherence to first order.
The final filled triangle and square
indicates that the ancillae are decoupled
while the data spin is observed.
} \label{fig:diagram}
\end{center} \end{figure}

\begin{figure}[H] \begin{center}
\begin{picture}(450,525) \put(-30,0){
\scalebox{1.0}[0.9]{\includegraphics{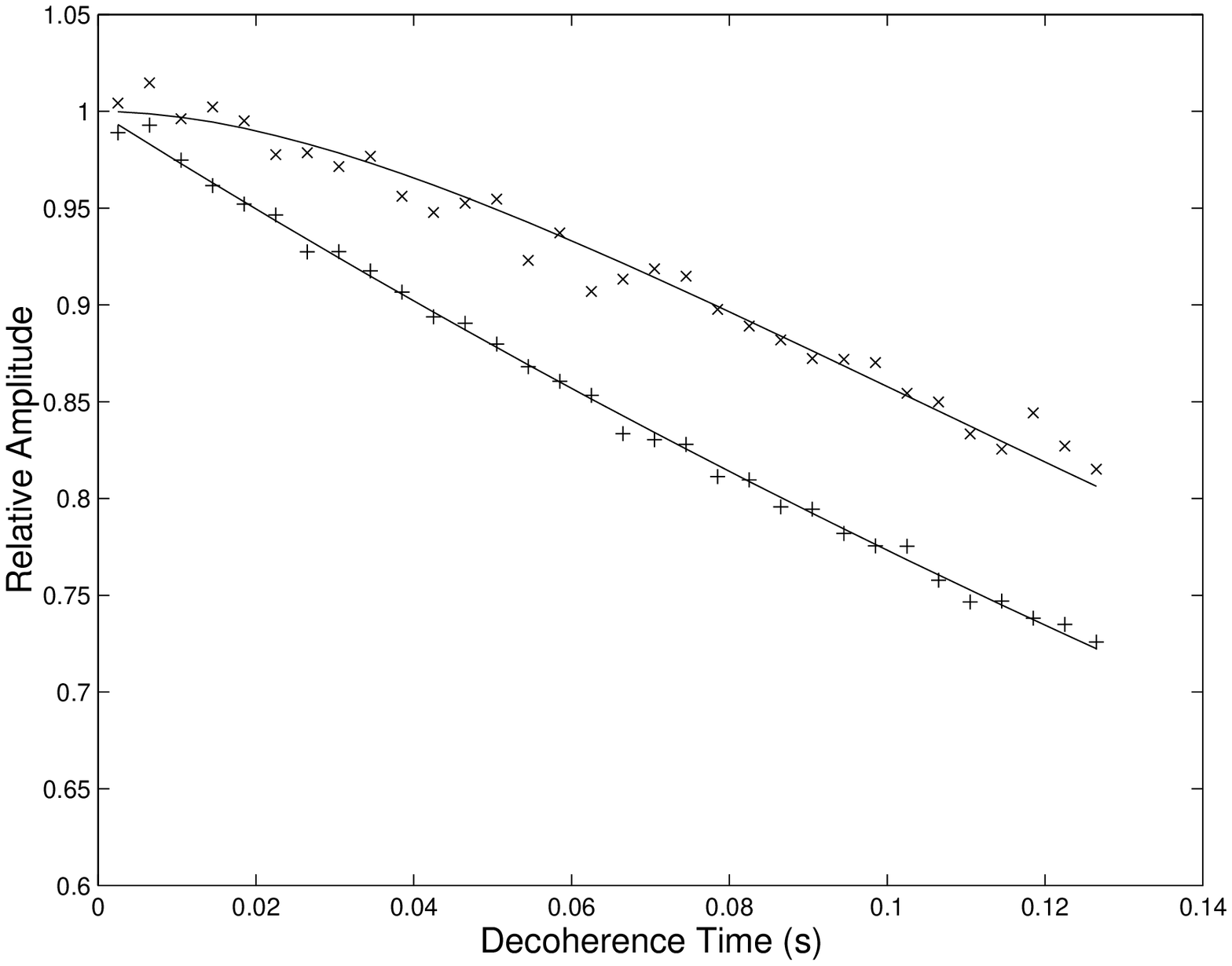}}
} \end{picture}
\bigskip \figcap{
Plots showing the experimental results for quantum
error correction applied to the $\IZ^1\EP^2\EP^3$
state (cf.\ Figure \ref{fig:diagram}).
The ``decoherence time'' is the
time allowed for diffusion in the
gradient-diffusion procedure (see text).
The amplitudes are of the peak due to the
data spin, following a $\pi/2$ readout pulse,
relative to its amplitude with no decoherence.
The ``$\LAB{x}$'' symbols mark the amplitudes
of the peak following the error correction
procedure at $32$ equally spaced decoherence times.
The ``$+$'' symbols mark the amplitudes of the
peak as a function of the decoherence time
when the gradient-diffusion procedure was
applied to the $\IX^1$ state with no error
correction at the same $32$ time points.
These measurements were averaged over $16$ repetitions
of each experiment (without any phase cycling).
The rate of decay of the peak $1/\tau = 2.5677$
$\mathrm{sec}^{-1}$ due to decoherence was obtained
by a linear least-squares fit to the logarithm of
the amplitudes of the peak from the $\IX^1$ state
(lower curve, correlation coefficient $-0.9987$),
after which the decay with error correction
was obtained from the theoretical relation
$(9 \exp( -t/\tau ) - \exp( -9t/\tau ))/8$
(upper curve, correlation coefficient $0.9873$).
} \label{fig:z_plot}
\end{center} \end{figure}

\begin{figure}[H] \begin{center}
\begin{picture}(450,525) \put(-30,0){
\scalebox{1.0}[0.9]{\includegraphics{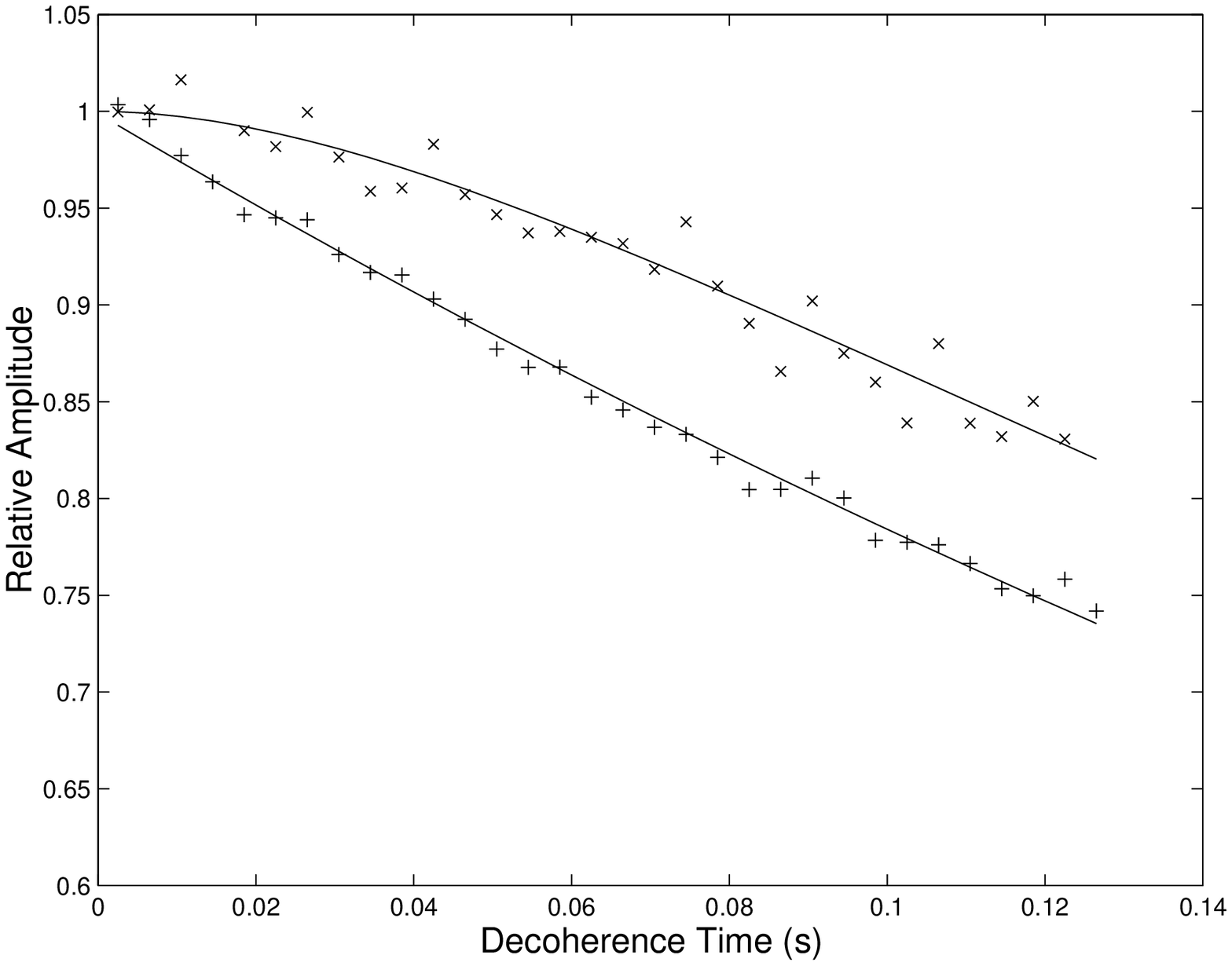}}
} \end{picture}
\bigskip \figcap{
Plots showing the experimental results for quantum
error correction applied to the $\IY^1\EP^2\EP^3$ state.
As in Figure \ref{fig:z_plot}, the ``$\LAB{x}$''
symbols mark the amplitudes of the peak due to the
data spin at $32$ equally spaced decoherence times
(note that the amplitudes at $t = 0.0145$ and
$t = 0.1265$ sec were treated as outlyers and omitted),
averaged over $16$ repetitions of each experiment.
The ``$+$'' symbols mark the amplitudes of the peak
from the $\IY^1$ state with no error correction,
averaged over $8$ repetitions of each experiment.
The rate of decay of the peak $1/\tau = 2.4200$
$\mathrm{sec}^{-1}$ due to decoherence was obtained
by a linear least-squares fit to the logarithm of
the amplitudes of the peak from the $\IY^1$ state
(lower curve, correlation coefficient $-0.9962$),
after which the decay with error correction
was predicted from the theoretical relation
$(9 \exp( -t/\tau ) - \exp( -9t/\tau ))/8$
(upper curve, correlation coefficient $0.9867$).
} \label{fig:y_plot}
\end{center} \end{figure}

\begin{figure}[H] \begin{center}
\begin{picture}(450,525) \put(-30,0){
\scalebox{1.0}[0.9]{\includegraphics{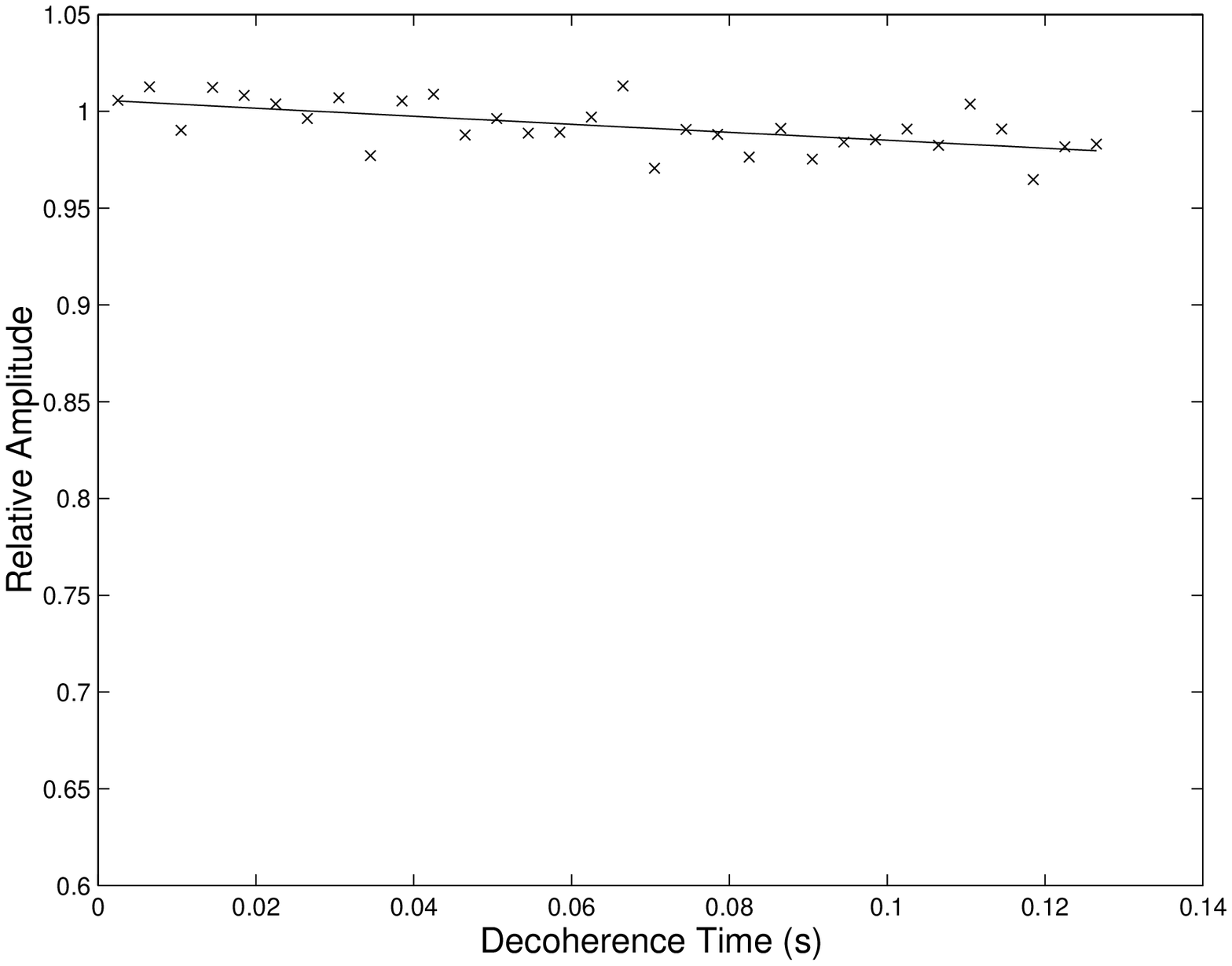}}
} \end{picture}
\bigskip \figcap{
Plots showing the experimental results for quantum
error correction applied to the $\IX^1\EP^2\EP^3$ state.
As in Figure \ref{fig:z_plot}, the ``$\LAB{x}$''
symbols mark the amplitudes of the peak due to the
data spin at $16$ equally spaced decoherence times,
averaged over $32$ repetitions of each experiment.
The decay rate of the peak $1/\tau = 0.2084$ due to
decoherence was obtained by a linear least-squares
fit to the logarithm of the amplitudes of the peak
(solid line, correlation coefficient $-0.6022$),
indicating that it is not decohered by gradient-diffusion
nor otherwise affected by the error correction procedure.
} \label{fig:x_plot}
\end{center} \end{figure}

\begin{figure}[H] \begin{center}
\begin{picture}(450,525) \put(-30,0){
\scalebox{1.0}[0.9]{\includegraphics{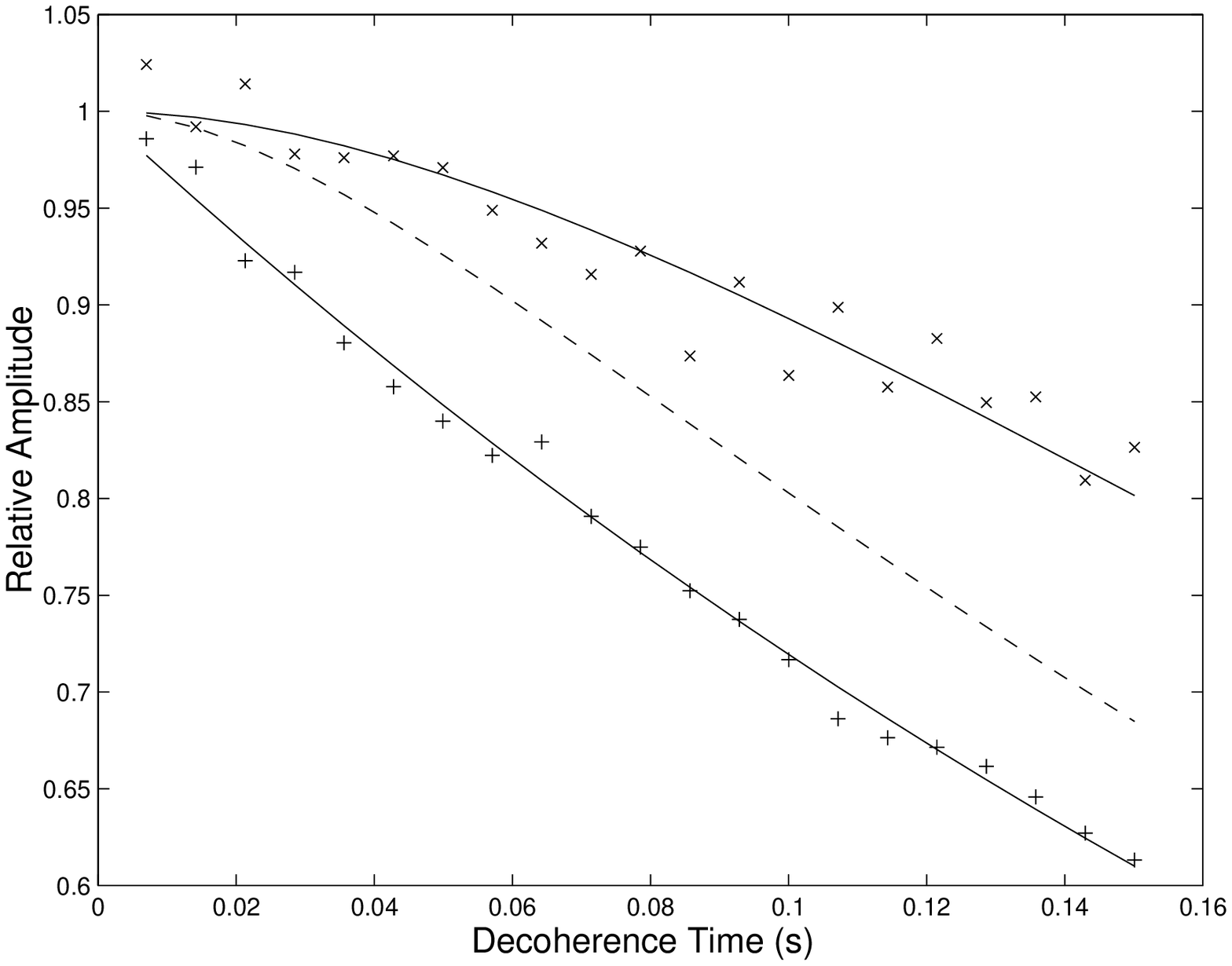}}
} \end{picture}
\bigskip \figcap{
Plots showing the experimental results for quantum
error correction applied to the $\IZ^1\EP^2\EP^3$ state,
with independent gradient-induced decoherence for each spin.
As in Figure \ref{fig:z_plot}, the ``$\LAB{x}$''
symbols mark the amplitudes of the peak due to the
data spin at $32$ equally spaced decoherence times
averaged over $16$ repetitions of each experiment.
The ``$+$'' symbols mark the amplitudes of the peak
from the $\IX^1\EP^2\EP^3$ state without encoding,
decoding or error correction,
averaged over $16$ repetitions of each experiment.
The rate of decay of the peak $1/\tau = 3.2931$
$\mathrm{sec}^{-1}$ due to decoherence was obtained by a
linear least-squares fit to the logarithm of the amplitudes of
the peak from the data spin as a function of decoherence time
(lower curve, correlation coefficient $-0.9970$),
after which the decay with error correction
was predicted from the theoretical relation
$(3 \exp( -t/\tau ) - \exp( -3t/\tau )) / 2$
(upper curve, correlation coefficient $0.9546$).
The decay curve predicted for totally correlated
decoherence as in Figure \ref{fig:z_plot} is
shown with a dashed line for comparison.
} \label{fig:u_plot}
\end{center} \end{figure}

This appendix shows that error correction will
not be able to correct the first-order decay of
the data spin in any diagonal mixed state for the
ancillae save for their pseudo-pure ground state,
unless the state of the data spin is taken into account.
Assuming that the ancillae are initially uncorrelated
with the data spin, such a state may be written as
\begin{equation}
\RHO ~=~ \RHO_\LAB{A}^1 \left(
\mu_{++} \EP^2 \EP^3 + \mu_{+-} \EP^2 \EM^3 +
\mu_{-+} \EM^2 \EP^3 + \mu_{--} \EM^2 \EM^3 \right)
\end{equation}
with nonnegative coefficients satisfying
$\mu_{++} + \mu_{+-} + \mu_{-+} + \mu_{--} = 1$.
For each term $\RHO_\LAB{A}^1\MAT E_{\epsilon^2}^2
\MAT E_{\epsilon^3}^3$ obtained upon expansion,
the only change in the preceding results (Eq.~(\ref{eq:final_result}))
is that the signs of the coefficients $F^2$ \&
$F^3$ are given by $\epsilon^2$ \& $\epsilon^3$,
so the partial trace after error correction is
\begin{equation}
\RHO_\LAB{E}^1 ~=~ \HALF + \nu_\LAB{x} 2\IX^1 +
(\nu_\LAB{y} 2\IY^1 + \nu_\LAB{z} 2\IZ^1)\, \Theta'(t) ~,
\end{equation}
where the partial trace is initially
\begin{equation}
\RHO_\LAB{A}^1 ~=~ \HALF + \nu_\LAB{x} 2\IX^1
+ \nu_\LAB{y} 2\IY^1 + \nu_\LAB{z} 2\IZ^1
\end{equation}
and now $\Theta'(t) =$
\begin{eqnarray}
&& \HALF \left( F^1 +
(\mu_{++} + \mu_{+-} - \mu_{-+} - \mu_{--}) F^2 +
(\mu_{++} - \mu_{+-} + \mu_{-+} - \mu_{--}) F^3
\right. \\ \nonumber && \left. \qquad -\,
(\mu_{++} - \mu_{+-} - \mu_{-+} + \mu_{--}) F^1F^2F^3 \, F^{123}
\right) ~.
\end{eqnarray}
The derivative at $t = 0$ is $\dot\Theta'(0) =$
\begin{equation}
\tfrac14 \left( (\mu_{++} - 1) c^{11}
- \mu_{+-} (c^{11} + 2c^{22}) - \mu_{-+} (c^{11} + 2c^{33})
+ \mu_{--} (c^{11} + 2c^{22} + 2c^{33}) \right) ~,
\end{equation}
and for $c^{11} \ne 0$ the only nonnegative solution to
this equation together with the normalization condition is
$\mu_{++} = 1$ and $\mu_{+-} = \mu_{-+} = \mu_{--} = 0$.
This proves that, in any such state, the three-bit
error correcting code (Fig.\ \ref{fig:diagram}) will
inhibit decoherence to first order only if the ancillae
are in a psuedo-pure state relative to the data spin.

More generally, consider a mixed state in which the ancillae are
diagonal but (classically) correlated with the data spin, i.e.
\begin{eqnarray}
\RHO &=& {\sum}_m\; \mu_m \,\RHO_m^1\,
\MAT E_{\epsilon_m^2}^2 \MAT E_{\epsilon_m^3}^3
\\ \nonumber &=&
\RHO_{++}^1 \EP^2 \EP^3 + \RHO_{+-}^1 \EP^2 \EM^3 +
\RHO_{-+}^1 \EM^2 \EP^3 + \RHO_{--}^1 \EM^2 \EM^3 ~,
\end{eqnarray}
where $\RHO_{++}^1 \equiv \sum_{\{m|\epsilon_m^2=\epsilon_m^3=1\}}
\mu_m \RHO_m^1$, etc., and $\sum_m \mu_m = 1$ ($\mu_m \ge0$).
Letting $\upsilon_{\epsilon^2\epsilon^3} \equiv \left\langle
\RHO_{\epsilon^2\epsilon^3}^1 2\IY^1 \right\rangle$
and $\zeta_{\epsilon^2\epsilon^3} \equiv \left\langle
\RHO_{\epsilon^2\epsilon^3}^1 2\IZ^1 \right\rangle$
be the $\LAB{y}$ and $\LAB{z}$ components of
the data spin's constituent density matrices
($\epsilon^2, \epsilon^3 \in \{ \pm1 \}$),
one can show that in this case the matrix
derivative at $t = 0$ vanishes if and only if
\begin{eqnarray}
\dot\Theta_{++} \upsilon_{++} + \dot\Theta_{+-} \upsilon_{+-} +
\dot\Theta_{-+} \upsilon_{-+} + \dot\Theta_{--} \upsilon_{--} &=& 0 \\
\dot\Theta_{++} \zeta_{++} + \dot\Theta_{+-} \zeta_{+-} + \nonumber
\dot\Theta_{-+} \zeta_{-+} + \dot\Theta_{--} \zeta_{--} &=& 0
\end{eqnarray}
where the $\Theta_{\epsilon^2\epsilon^3}$ are
obtained by changing the signs of $F^2$ and $F^3$
in our expression for $\Theta \equiv \Theta_{++}$.
Since $\dot\Theta_{++}(0) = 0$, this is a system
of two linear equations in the three unknowns
$-2\dot F^1 = c^{11}$, $-2\dot F^2 = c^{22}$ and
$-2\dot F^3 = c^{33}$, which is generally solvable.
Nevertheless, the coefficients in these equations
depend on the state of the data to be protected
and its correlations with the ancillae,
which makes it impossible to use the three-bit
code to protect unknown data in this case.

\bibliographystyle{unsrt}
\bibliography{../../math,../../csci,../../nmr,../../phys,../../self}

\pagebreak \mylistoffigs

\pagebreak[4]

\pagebreak[4]

\pagebreak[4]

\pagebreak[4]

\pagebreak[4]

\pagebreak[4]

\end{document}